%% file: main.tex
\documentclass[a4paper,
onecolumn, published, accepted=2019-12-11,
superscriptaddress,11pt]{quantumarticle}
\pdfoutput=1
\usepackage[utf8]{inputenc}
\usepackage[english]{babel}
\usepackage[T1]{fontenc}
\usepackage{amsmath}
\usepackage{hyperref}

\usepackage{tikz}
\usepackage{lipsum}

\makeatletter 
 \hypersetup{pdftitle = {pdf title},
	     pdfauthor = {Daniel Miller},
	     pdfsubject = {Quantum Communication},
	     pdfkeywords = {PLOB, repeaterless bound,  
			    genuine, error-corrected, one-way, quantum repeater,
			    quantum-repeater gain,
			    quantum entanglement, quantum internet,
			    error analysis,   
			    pure-loss channel, bosonic channel,
			    multimode, Fock, qudit, 
			    generalized Pauli, Weyl, Heisenberg,
			    quantum polynomial code, optimal distance, optimal spacing
			    }
	    }
\makeatother

\usepackage{cite}   
\usepackage{braket}
\usepackage[normalem]{ulem}   
\usepackage{amsmath,amssymb}  
\usepackage{stmaryrd} 

\newcommand{\CZ}{\textsc{\MakeLowercase{CZ}} }
\newcommand{\ZDZ}{\mathbb{Z}/D\mathbb{Z}}
\newcommand{\e}{\mathrm{e}}
\newcommand{\round}{\mathrm{rd}}
\newcommand{\wt}{\mathrm{wt}}

\newcommand{\lbrep}{{B}^\downarrow_\mathrm{rep} }
\newcommand{\ubPLOB}{{B}^\uparrow_\mathrm{PLOB} }

  \usetikzlibrary{shapes,arrows,spy,positioning,snakes}

\begin{document}
 
\title{Parameter regimes for surpassing the PLOB bound with error-corrected qudit repeaters}
\date{\today}		
\author{Daniel Miller} \orcid{0000-0003-2100-5612} \email{daniel.miller@hhu.de}
\author{Timo Holz} 
\author{Hermann Kampermann} \orcid{0000-0002-0659-6699}
\author{Dagmar Bru\ss} 
\affiliation{Institut f\"ur Theoretische  Physik III, Heinrich-Heine-Universit\"at D\"usseldorf, D-40225 D\"usseldorf, Germany}

\maketitle


\begin{abstract} 
\input{abstract}
\end{abstract} 

\tableofcontents

\section{Introduction}

\input{1Introduction.tex} 

\section{Identification of genuine quantum repeaters}\label{sec:methods}		
\input{2Setup.tex}

\newpage

\section{Parameter regimes for genuine quantum repeaters}\label{sec:results}
\input{3Results.tex}

\section{Conclusion and Outlook}\label{sec:conclusion}
\input{4Conclusion.tex}

\end{document}

%% file: abstract.tex
 
A potential quantum internet would open up the possibility of realizing numerous new applications, including provably secure communication. 
Since losses of photons limit long-distance, direct quantum communication and wide\-spread quantum networks, quantum repeaters are needed. 
The so-called PLOB-repeaterless bound [Pirandola et \emph{al.}, Nat. Commun. \textbf{8,} 15043 (2017)] 
is a fundamental limit on the quantum capacity of direct quantum communication.   
Here, we analytically derive the quantum-repeater gain for error-corrected, one-way quantum
repeaters based on higher-dimensional qudits for two different physical encodings:
Fock and multimode qudits.
We identify parameter regimes in which such quantum repeaters can surpass the PLOB-repeaterless bound and systematically analyze how typical parameters manifest themselves in the quantum-repeater gain.
This benchmarking provides a guideline for the implementation of error-corrected qudit repeaters.
 

%% file: 1Introduction.tex
The  prospect of an eventual world-spanning quantum internet 
motivates tremendous interest and investments~\cite{RBTC17, ABBCEEEGGJKLRSTWWW18, WEH18}. 
A quantum internet offers---among an increasing number of other applications \cite{WEH18, GLM01, ChrWeh05, GJC12, KBdGL18}---the possibility of quantum key distribution (QKD),
a cryptographic procedure whose security is not based on computational hardness assumptions but on the laws of quantum mechanics~\cite{BB84, Ekert91, Bruss98}.
Although state-of-the-art experiments can perform fiber-based direct-transmission QKD across a few hundred kilometers~\cite{BBRVACPGBLNMZ18}, they face fundamental limitations~\cite{TGW14, ChrMH07, PLOB17}.
The so-called PLOB-repeaterless bound (named after Pirandola, Laurenza, Ottaviani and Banchi) states that 
the quantum capacity of a fiber directly connecting two parties is exponentially suppressed in their distance~\cite{PLOB17}.
As the quantum capacity is closely related to the amount of transmissible quantum information, direct transmission channels are not well suited for long distance quantum connections.  
To overcome these limitations, quantum repeaters have been proposed~\cite{BDCZ98, vLLSYNMY06, JTNMVL09, FWHLVH10, MKLLJ14, MLKLLJ16}.
They shorten the distance of direct transmissions by introducing intermediate repeater stations such that losses and errors can be tackled using entanglement heralding, quantum memories, entanglement distillation, or quantum error-correcting codes (QECCs)~\cite{MLKLLJ16}.     
Recent investigations have shown that quantum repeaters based on currently available technology have the potential to surpass the PLOB-repeaterless bound, even with a single intermediate repeater station~\cite{LJKL16, RGRKCVRSE18, RYGRHHWE18,LYDS18,  CAL18, GraCur19}.
Laboratory experiments have been reported which prove that it is in principle possible to surpass the PLOB-repeaterless bound  
over distances of tens and hundreds of kilometers~\cite{MPRLDYS19, WHYLCCZGZ19, ZHCQL19}.
 
For world-spanning quantum communication, error-corrected, one-way~\cite{JTNMVL09, FWHLVH10, MKLLJ14} (also known as third generation~\cite{MLKLLJ16}) quantum repeaters are promising candidates. 
Since the implementation of such quantum repeaters will be demanding and expensive, it is crucial to identify under which circumstances they can be superior to direct quantum communication. 
{In Refs.~\cite{EwLoock17, SchLoock19}, this is done for qubit based quantum repeaters employing static linear optics.}
Here, we address this problem in the case of error-corrected, one-way quantum repeaters based on qudits (discrete variable quantum systems of dimension $D\ge2$), as such higher-dimensional qudits offer the advantage that more noise can be tolerated before entanglement is lost~\cite{EBBBKSFMGUH19}. 
(See Refs.~\cite{MZLWJ17, MZLJ18, MHKB18, BvL19} for previous investigations in quantum repeaters based on qudits.)
To conclude that a quantum repeater can overcome the PLOB-repeaterless bound, it is instrumental to find a lower bound on the achievable quantum capacity of quantum repeaters.
In previous approaches~\cite{LJKL16, RGRKCVRSE18, RYGRHHWE18, CAL18, GraCur19, LYDS18}, this figure of merit 
usually was given by the secret key rate achievable with a specific protocol; see also Refs.~\cite{ABBKvLB13, SHB14, EKB16_analysis, EKB16_networks, EKB16_router}
for earlier investigations in secret key rates of quantum repeaters. 
In this paper, we use a different approach by exploiting that the quantum capacity of an error-corrected quantum repeater can be lower bounded by $\log_2(D) - H(P)$, where $H(P)$ is the Shannon entropy of the error probability distribution $P$ of the state distributed by the repeater~\cite{PLOB17, VW03}.   
There are numerous parameters influencing the performance of quantum repeaters, e.g., total distance, number of intermediate repeater stations, various error rates and choice of a QECC.
In Ref.~\cite{MHKB18}, we derived an expression of the error probability distribution $P$ in terms of these parameters. 
Here, we identify and discuss parameter regimes in which error-corrected, one-way quantum repeaters based on qudits can beat the PLOB-repeaterless bound. 
 
This paper is structured as follows. In Sec.~\ref{sec:methods}, we explain our method to assess the quantum capacity of quantum repeaters. In Sec.~\ref{sec:results}, we identify parameter regions where error-corrected qudit repeaters can surpass the PLOB-repeaterless bound. Finally, in Sec.~\ref{sec:conclusion} we conclude and give an outlook on possible future work.

%% file: 2Setup.tex
Consider two remote parties called Alice and Bob. 
A quantum channel $\mathcal{E}$ from Alice to Bob is a completely positive, trace-preserving map from the space of density operators on Alice's Hilbert space to that of Bob.
The (two-way) \emph{quantum capacity}, $\mathcal{C}(\mathcal{E})$, quantifies how much quantum information Alice can transmit asymptotically to Bob through  $\mathcal{E}$ using adaptive local operations and classical communications. 
See Ref.~\cite{PLOB17} for the formal definition. 
We call such a quantum channel a \emph{genuine quantum repeater} if it has a quantum capacity that is larger than that of any direct transmission. 
{In Sec.~\ref{sec:PLOB_bound}, we explain how this characterization can be assessed via the multimode PLOB-repeaterless bound.
In Sec.~\ref{sec:bosonic_qudits}, we propose two physical encodings of qudits into photons and
provide error models for the transmission of these bosonic qudits. }
In Sec.~\ref{sec:abstract_qudits}, we recall a more abstract description of qudits which we will use 
throughout this paper.
In Sec.~\ref{sec:repeater_protocol}, we present the here-considered protocol for an error-corrected qudit repeater and define its \emph{quantum-repeater gain}.
If this figure of merit is positive, a  genuine quantum repeater is identified.

\subsection{{The PLOB-repeaterless bound}} 
\label{sec:PLOB_bound}

Consider a photonic mode with a bosonic creation operator $b^\dagger$.
The \emph{pure-loss channel} $\mathcal{E}^{(\eta)}_\mathrm{loss}: 
 b^\dagger \mapsto \sqrt{\eta}\, b^\dagger + \sqrt{1-\eta}\, b_\mathrm{E}^\dagger, $ 
mixes such a mode with a vacuum mode via a beam splitter with transmissivity $\eta$, 
where $ b_\mathrm{E}^\dagger$ is the creation operator of an environmental bosonic mode initialized in the zero-photon state $\ket{0}_\mathrm{E}$~\cite{BrauLoock05, WPGCRSL12}. 
The \emph{PLOB-repeaterless bound} states that the quantum capacity of every quantum channel $\mathcal{E}$ is limited by that of $\mathcal{E}^{(\eta)}_\mathrm{loss}$, 
\begin{align}\label{eq:PLOB}  
 \mathcal{C}(\mathcal{E}) \le \mathcal{C}(\mathcal{E}^{(\eta)}_\mathrm{loss}) = - \log_2(1-\eta),
\end{align}
provided there exists a decomposition of the form $\mathcal{E}=\mathcal{E}_\mathrm{B}\circ \mathcal{E}^{(\eta)}_\mathrm{loss} \circ \mathcal{E}_\mathrm{A}$ 
for some quantum channels $\mathcal{E}_\mathrm{A}$ and $\mathcal{E}_\mathrm{B}$~\cite{PLOB17}. 
As this type of decomposition is typical for direct quantum communication scenarios, this bound is fundamental.
Moreover, such decompositions are known for all Gaussian channels~\cite{WPGCRSL12, PLOB17}.
In order to overcome this limit on direct quantum communication, one could employ $M>1$ bosonic modes in parallel and perform measurements at repeater stations in between Alice and Bob. 
Since the quantum capacity of the pure-loss channel is additive, the quantum capacity $\mathcal{C}(\mathcal{E}_\mathrm{rep})$ of a genuine quantum repeater of this form has to beat the  \emph{multimode PLOB-repeaterless bound},
\begin{align}\label{eq:repeaterless_PLOB_bound}
 \mathcal{C}(\mathcal{E}_\mathrm{direct}) \le - M \times \log_2(1-\eta).
\end{align}
This is the ultimate limit of direct quantum communication using $M$ bosonic modes through free space or an optical fiber with transmissivity $\eta$.  
At high loss $\eta\approx0$, the multimode PLOB-repeaterless bound scales linearly in the transmissivity according to {$-M \times \log_2(1-\eta)\approx1.44M\eta$, i.e., it is exponentially suppressed in the distance~\cite{PLOB17}.

\subsection{{Bosonic qudits}}
\label{sec:bosonic_qudits}
{In order to investigate under which circumstances a quantum repeater has the potential to overcome the PLOB-repeaterless bound, it is necessary to specify the physical encoding of quantum information into bosonic modes. We consider two types of bosonic qudits: one-mode Fock qudits and multimode qudits. Here, we specify how the pure-loss channel acts on these encodings.}

The pure-loss channel transforms a pure Fock number state 
$\ket{k}= \frac{1}{\sqrt{k!}}(b^\dagger)^ k \ket{0}$ into
\begin{align}\label{eq:lossy_on_fock}
 \mathcal{E}^{(\eta)}_\mathrm{loss} \left(\ket{k}\bra{k} \right) = 
 \sum_{j=0}^k {k \choose j} \eta^j(1-\eta)^{k-j} \, \ket{j} \bra{j}.
\end{align} 
Let $\mathcal E^{(\eta)}_{\mathrm{Fock};D}$ denote the $D$-dimensional restriction of the pure-loss channel to inputs with $k\le D-1$ photons. 
This channel possesses a decomposition 
$\mathcal E^{(\eta)}_{\mathrm{Fock};D}=\mathcal{E}_\mathrm{CV\rightarrow DV}\circ \mathcal{E}^{(\eta)}_\mathrm{loss} \circ \mathcal{E}_\mathrm{DV\rightarrow CV}$, where $\mathcal{E}_\mathrm{DV\rightarrow CV}$ is the inclusion map from $\mathbb{C}^D$ to the single mode Fock space, i.e.,  $\mathcal{E}_\mathrm{DV\rightarrow CV}$ sends a computational basis state to the state with the corresponding photon number,  and similarly for $\mathcal{E}_\mathrm{CV\rightarrow DV}$, cf. Ref.~\cite{PLOB17}. 
Thus, the PLOB-repeaterless bound yields
\begin{align} 
 \mathcal C\left( \mathcal{E}^{(\eta)}_{\mathrm{Fock},D} \right) \le 
 \mathcal{C}(\mathcal{E}^{(\eta)}_\mathrm{loss}) =	 - \log_2(1-\eta),
\end{align}
where equality is reached in the limit $D\rightarrow \infty$. 
However, if $D$ is finite, the inequality is strict, as not the full potential of the pure-loss channel is exploited. 
%
 
Instead of encoding a $D$-dimensional qudit into the Fock basis, one could also use time-bin encoding~\cite{BGTZ99, dRMZG02, ZZHLVLRBMGNMSZWEWSW15, MMBVNMSG18}, temporal modes (TM)~\cite{BRSR15, ADASBRTHS18} or modes of orbital angular momentum (OAM)~\cite{ABSW92, CaPiBa06, PKFRZ13}.
For any of these implementations---to which we will from now on refer to as \emph{multimode encoding}---the computational basis states are given by a single photon in one of $D$ modes, i.e.,
\begin{align}
 \ket{m} := b^\dagger_m \ket{\mathrm{vac}},
\end{align}
where $b^\dagger_m$ is the creation operator of a bosonic mode labeled by $m\in\{0,\ldots,D-1\}$ and $\ket{\mathrm{vac}}=\ket{0,\ldots,0}$ is the vacuum state of all $D$ modes.
Sending {a multimode qudit in state $\rho$} through a beam splitter {(with zero environmental photons)} gives rise to the $D$-dimensional erasure channel,
\begin{align}\label{eq:erase}
 \mathcal{E}^{(\eta)}_{\mathrm{erase};D} :  \rho \longmapsto \eta \, \rho + (1-\eta)\, \ket{\mathrm{vac}}\bra{\mathrm{vac}}.
\end{align}
Its quantum capacity is known to be $\mathcal{C}\left( \mathcal{E}^{(\eta)}_{\mathrm{erase};D}\right)  = \eta \log_2(D)$ \cite{PLOB17}. {This is smaller than $\mathcal{C} \left( (\mathcal{E}^{(\eta)}_\mathrm{loss})^{\otimes D} \right) = -D\times \log_2(1-\eta)$ because $\mathcal{E}^{(\eta)}_{\mathrm{erase};D} $ is only defined on the domain of one-photon excitations, while for  $D$ pure-loss channels arbitrary superpositions of $D$-mode Fock states are valid inputs.
}

\subsection{Abstract description of qudits}
 \label{sec:abstract_qudits}
 
Formally, a qudit is a quantum system with a Hilbert space of dimension $D\ge2$.  
We label its computational basis states $\ket{j}$
by elements $j$ in
$\ZDZ=\{0,1,\ldots,D-1\}$, the ring of integers modulo $D$. 
For example,  
\begin{align}
\ket{+}:=\frac{1}{\sqrt{D}}\sum_{j\in\ZDZ} \ket{j},
\end{align}
is the equally weighted superposition of all computational basis states.
Up to a global phase, the generalized Pauli-operators of a qudit are products of  
\begin{align}\label{eq:PauliXZ}
 X := \sum_{k\in \ZDZ} \ket{k+1}\bra{k}  
 \hspace{2em} \text{and} \hspace{1em}
  Z := \sum_{k\in \ZDZ} \omega^k \ket{k}\bra{k},
\end{align} 
where $\omega:=\e^{2\pi i / D}$, i.e., the generalized Pauli-operators are of the form
\begin{align}\label{eq:pauli}
X^rZ^s  =\sum_{k\in \ZDZ} \omega^{ks} \ket{k+r}\bra{k},
\end{align} 
where $r,s\in \ZDZ$. 
They constitute a basis of the vector space of complex $D\times D$ matrices~\cite{Knill93, Gottesman99}.
The generalized Pauli-error channel, given an error probability distribution ${P=(p_{r,s})_{r,s\in \ZDZ}}$ with $\sum_{r,s}p_{r,s}=1$, is defined as the quantum channel,
\begin{align}\label{eq:pauli_channel}
\mathcal{E}_P:\rho \longmapsto \sum_{r,s \in \ZDZ} p_{r,s} (X^r Z^s) \rho (X^r Z^s )^\dagger,
\end{align} 
with Kraus operators $\sqrt{p_{r,s}}X^{r}Z^{s}$.
It corresponds to the random application of a Pauli-operator $X^rZ^s$ to the state $\rho$ with probability $p_{r,s}$.   
The depolarizing channel,
\begin{align}\label{eq:depol}
\mathcal{E}_{\mathrm{depol};D}^{(1-f)}: \rho \longmapsto (1-f)\rho + f \frac{\mathbbm{1}}{D},
\end{align}
is an example of a generalized Pauli-error channel where the trivial error $X^0Z^0=\mathbbm{1}$ occurs with probability $p_{0,0}= 1 -f+ f/D^{2}$ and any other error occurs with probability $f/D^{2}$~\cite{MHKB18}.
The controlled-$Z$ gate, 
\begin{align}
  \CZ := \sum_{k\in \ZDZ} \ket{k}\bra{k} \otimes Z^k
\end{align}
is a two-qudit Clifford gate which can be used to produce the maximally-entangled state,
\begin{align}\label{eq:entangled_state}
  \ket{\Phi}:= \CZ \ket{+}^{\otimes 2} =  \frac{1}{D}\sum_{j,k\in\ZDZ} \omega^{jk}\ket{j}\otimes \ket{k},
\end{align}
from two copies of the $\ket{+}$ state.

\subsection{Error-corrected qudit repeaters and the quantum-repeater gain }
\label{sec:repeater_protocol}

Assume that Alice and Bob make use of the one-way quantum repeater protocol described in the caption of Fig.~\ref{fig:repeater}.
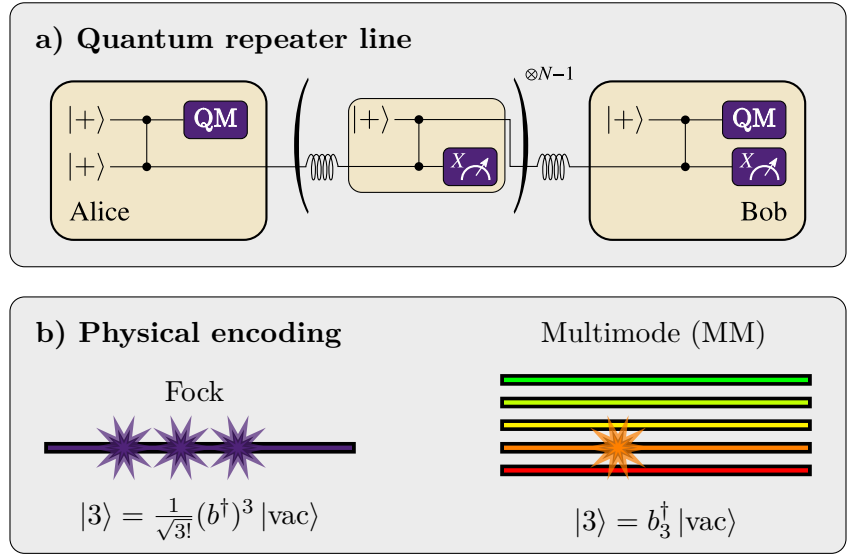
\begin{figure}[t]  
 \definecolor{Qlila} {rgb}{0.3086,0.1367,0.4688}   
 \definecolor{Qbeige}{rgb}{0.91,0.85,0.72}   
\begin{center}  
   \begin{tikzpicture}
   
  \fill[gray, opacity=0.15, rounded corners=.5em] (2.5,2.5) rectangle (-8.5,6); 
  \draw[rounded corners=.5em] (2.5,2.5) rectangle (-8.5,6);   
   \draw (-8.3, 5.5) node[anchor=west, text=black]{\textbf{a) Quantum repeater line}};  
   
    \node (RO) at (-3, 4) {\includegraphics[width= 26em]{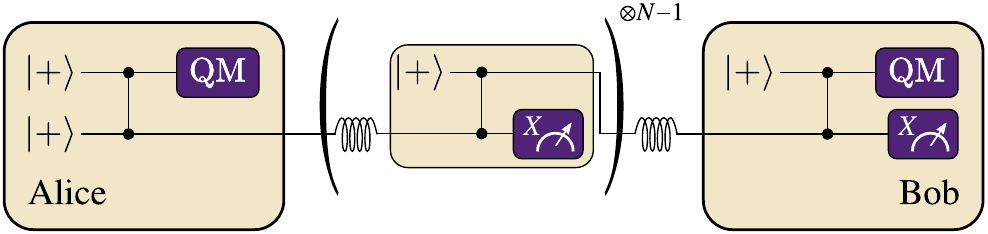}};

  \fill[gray, opacity=0.15, rounded corners=.5em] (2.5,-1.3) rectangle (-8.5,2.1); 
  \draw[rounded corners=.5em] (2.5,-1.3) rectangle (-8.5,2.1);  
   
   \draw (-8.3, 1.6) node[anchor=west, text=black]{\textbf{b) Physical encoding}};

   \draw (-1.65, 1.6) node[anchor=west,text=black]{Multimode (MM)};  
   \draw (-6.6, 0.85) node[anchor=west,text=black]{Fock};


    \draw (0, -.85) node[text=black]{$\ket{3} = { b_3 ^\dagger} \ket{\mathrm{vac}}$}; 
    
    \draw[black, line width=.4em] (-2.05,-0.2) -- (2.05,-0.2);    
    \draw[black, line width=.4em] (-2.05,0.1) -- (2.05,0.1);
    \draw[black, line width=.4em] (-2.05,0.40) -- (2.05,0.40); 
    \draw[black, line width=.4em] (-2.05,0.7) -- (2.05,0.7); 
    \draw[black, line width=.4em] (-2.05,1.00) -- (2.05,1.00);     
     
    \draw[opacity = 1.75, red ,    line width=.15em] (-2,-0.2) -- (2,-0.2);    
    \draw[opacity = 1.75, orange , line width=.15em] (-2,.1)  -- (2,.1);
    \draw[opacity = 1.75, yellow , line width=.15em] (-2,0.4) -- (2,0.4); 
    \draw[opacity = 1.75, lime, line width=.15em]  (-2.0,0.7) -- (2.0,0.7); 
    \draw[opacity = 1.00, green, line width=.15em] (-2.0,1.00) -- (2.0,1.00);

    \node[darkgray , fill,     star, star points=10, star point ratio=.4]at (-0.5,0.1) {.}; 
    \node[orange, fill,              star, star points=10, star point ratio=.3] at (-0.5,0.1) {};
    \node[opacity=.7, orange , fill, star, star points=10, star point ratio=.4] at (-0.5,0.1) {\hspace{1em} };

     \draw (-6, -0.85) node[text=black]{$\ket{3} = \frac{1}{\sqrt{3!}} {(b ^\dagger)^3 } \ket{\mathrm{vac}}$}; 
     
    \draw[black, line width=.4em] (-3.95,0.1) -- (-8.05,0.1);     
    
    \draw[opacity = 1.75, Qlila ,    line width=.15em] (-4,0.1) -- (-8,0.1);

    \node[fill,                     star, star points=10, star point ratio=.4] at (-5.5,0.1) {.}; 
    \node[Qlila, fill,              star, star points=10, star point ratio=.3] at (-5.5,0.1) {};
    \node[opacity=.7, Qlila , fill, star, star points=10, star point ratio=.4] at (-5.5,0.1) {\hspace{1em} }; 
     
    \node[fill,                     star, star points=10, star point ratio=.4] at (-6.25,0.1) {.}; 
    \node[Qlila, fill,              star, star points=10, star point ratio=.3] at (-6.25,0.1) {};
    \node[opacity=.7, Qlila , fill, star, star points=10, star point ratio=.4] at (-6.25,0.1) {\hspace{1em} }; 
    
    \node[fill,                     star, star points=10, star point ratio=.4] at (-7.,0.1) {.}; 
    \node[Qlila, fill,              star, star points=10, star point ratio=.3] at (-7.,0.1) {};
    \node[opacity=.7, Qlila , fill, star, star points=10, star point ratio=.4] at (-7.,0.1) {\hspace{1em} };     
   \end{tikzpicture}
\end{center}

\caption {\textbf{a)} An error-corrected, one-way qudit quantum repeater line with $N-1$ intermediate repeater stations~\cite{EKB16_analysis, EKB16_networks, EKB16_router, MHKB18}. 
Alice produces the two-qudit state $\ket{\Phi}=\CZ\ket{+}^{\otimes2}$, stores one qudit into a quantum memory (QM), and sends the other qudit to the first intermediate repeater station.
At every repeater station, the incoming qudit is entangled via a $\CZ$ gate with a new qudit prepared in the $\ket{+}$ state.
Then, the previous qudit is measured in the $X$ basis and the other qudit is sent to the next repeater station.
After $N$ transmissions, Bob receives the last qudit, entangles it with his own $\ket{+}$ state, measures it, and stores the remaining qudit in his own QM. 
As a result, Alice and Bob have stored an entangled qudit pair in their QMs.
The protocol takes place on a logical level where each logical qudit consists of $n$ physical qudits.
\textbf{b)} Visualization of two different encoding methods into photons. 
}
\label{fig:repeater} 
\end{figure}  
In the ideal case, they obtain a maximally entangled state $B^\dagger\ket{\Phi}$ in their quantum memories (QMs) which only differs from $\ket{\Phi}$ of Eq.~\eqref{eq:entangled_state} by the application of the byproduct operator $B=X^{c_\mathrm{even}}Z^{c_\mathrm{odd}}$ to Bob's qudits, where the number of elementary links $N$ is even.
The exponents
\begin{align}\label{eq:byproduct}
c_\mathrm{even}:= \sum_{i=1}^{N/2} (-1)^{i} c_{2i}
\hspace{2em} \text{and} \hspace{1em}
c_\mathrm{odd} := \sum_{i=1}^{N/2} (-1)^{i+1} c_{N+1-2i},
\end{align}  
are computed from the measurement outcomes $c_i$ at the $i$-th repeater station,
i.e., $c_\mathrm{even}$ depends on the measurement outcomes of even-numbered repeater stations and likewise for $c_\mathrm{odd}$.
See Ref.~\cite{MHKB18} for more details.
  
To overcome the limit of direct quantum communication, the quantum repeater employs logical qudits which are encoded using an $\llbracket n,1,d \rrbracket_D$ QECC such that each qudit is replaced by $n$ physical qudits of dimension $D$. 
The code distance $d$ determines the number $t$ of correctable errors by the QECC according to $t=\lfloor (d-1)/2\rfloor$, see Ref.~\cite{QEC} for its formal definition.  
Note that QECCs do not exist for all code parameters, e.g., all QECCs fulfill the quantum singleton bound $2d-1\le n$. 
However, if $D$ is a prime number and $d\le(D-1)/2$, an explicit construction of $\llbracket 2d-1,1,d \rrbracket_D$ QECCs saturating the quantum singleton bound is known in the form of quantum polynomial codes~\cite{GL99, ABO08, KKKS06, Cross08}. 
We focus on this encoding because quantum polynomial codes can give an advantage over other QECCs for quantum repeaters~\cite{MZLWJ17, MZLJ18}.

At every repeater station, the (logical) $X$ measurement is performed as follows.
All physical qudits are measured individually in the eigenbasis of the $X$ operator. 
If the number of erroneous measurement results is smaller than or equal to $t$,  
the error pattern can be corrected successfully. 
Otherwise, {the correction attempt leads to a logical error. For the considered QECCs, it is infeasible to determine the success probability for this strategy. 
Thus, we bound the capacity of the repeater by using a random logical error (instead of the actual logical error) whenever the number of physical errors is larger than~$t$. 
Note that this simplification does not overestimate the capacity of the quantum repeater. We will make this statement more precise in Eq.~\eqref{eq:lbrep}.
For this strategy, it} turns out~\cite{MHKB18} that all repeater stations (except for the first) have the same probability   $p^\mathrm{rep}_\mathrm{succ}$ of a successful correction.
For the first repeater station, where fewer error sources contribute, we denote this probability  
by $p^{1.\mathrm{rep}}_\mathrm{succ}$. 
To simplify the error analysis, we furthermore assume that Bob performs a (perfect) round of stabilizer measurements.
As quantum polynomial codes belong to the class of Calderbank-Shor-Steane codes~\cite{Cross08}, Bob can independently correct $X$ and $Z$ errors.
Again, if the number of $X$ ($Z$) errors exceeds $t$, Bob guesses a recovery operation. 
Otherwise, with probability $p_\mathrm{succ}^{\mathrm{Bob},X}$ ($p_\mathrm{succ}^{\mathrm{Bob},Z}$),  
he can successfully reveal the error pattern and applies the appropriate recovery operation.
In the final step of the entanglement swapping protocol, Bob has to apply the byproduct operator $B=X^{c_\mathrm{even}}Z^{c_\mathrm{odd}}$. 
In doing so, errors on the measurement results of the even-numbered repeater stations propagate to $X$ errors on Bob's qudit which gives rise to an $X$ dephasing channel, $\rho \mapsto p_\mathrm{succ}^{\mathrm{rep}}\rho + \frac{1-p_\mathrm{succ}^{\mathrm{rep}} }{D}\sum_{r\in \ZDZ} X^r \rho X^{-r}$.
Likewise, wrong measurement results at odd-numbered repeater stations induce $Z$ errors on Bob.
%
In conclusion, the overall error statistics are of the form
$P=(p^\mathrm{fin}_{r,s})_{r,s\in\ZDZ} = (p^{\mathrm{fin},X}_rp^{\mathrm{fin},Z}_s)_{r,s\in\ZDZ}$, where
\begin{align}\label{eq:p_fin_x}
 p^{\mathrm{fin},X}_0 &= 
                         \frac{1}{D}\left[1+(D-1)(p^\mathrm{rep}_\mathrm{succ})^\frac{N}{2}p^{\mathrm{Bob},X}_\mathrm{succ}  \right] \text{ and}
\\\nonumber
p^{\mathrm{fin},X}_{r\neq 0} &=
                         \frac{1}{D}\left[1-(p^\mathrm{rep}_\mathrm{succ})^\frac{N}{2}p^{\mathrm{Bob},X}_\mathrm{succ} \right] 
\end{align}
are the probabilities of $X$ errors on Bob's qudit, and 
\begin{align} \label{eq:p_fin_z}
 p^{\mathrm{fin},Z}_0 &= 
                         \frac{1}{D}\left[1+(D-1) p^\mathrm{1.rep}_\mathrm{succ}(p^\mathrm{rep}_\mathrm{succ})^{\frac{N}{2}-1}p^{\mathrm{Bob},Z}_\mathrm{succ} \right] \text{ and}
                      \\ \nonumber
p^{\mathrm{fin},Z}_{s\neq 0} &=
			\frac{1}{D}\left[1-p^\mathrm{1.rep}_\mathrm{succ}(p^\mathrm{rep}_\mathrm{succ})^{\frac{N}{2}-1}p^{\mathrm{Bob},Z}_\mathrm{succ} \right]
\end{align}
are the probabilities of $Z$ errors.
We will explain in Sec.~\ref{sec:stats} how these success probabilities depend on the physical error rates.

The erroneous state distributed by the quantum repeater is $\rho = \mathcal{E}_P( \Phi)$, where  $\Phi=\ket{\Phi}\bra{\Phi}$ is the projector onto the maximally entangled state defined in Eq.~\eqref{eq:entangled_state}, and $\mathcal{E}_P$ is the generalized Pauli-error channel, recall Eq.~\eqref{eq:pauli_channel}, corresponding to the error distribution $P$ acting on Bob's qudit. 
The quantum capacity of a generalized Pauli-error channel $\mathcal{C}(\mathcal{E}_P)$ can be lower bounded by 
\begin{align} \label{eq:lbrep}
 \mathcal{C}(\mathcal{E}_\mathrm{rep}) = \mathcal{C}(\mathcal{E}_P)  \ge  \max\{ 0,\,  \log_2(D) - H(P) \} =: \lbrep,
\end{align}
see suppl. of Ref.~\cite{PLOB17}. 
{As mentioned before, guessing logical errors leads to an overestimation of $H(P)$, thus, even for the implementable strategy, $\lbrep$ is still a lower bound on the capacity of the quantum repeater.} 
Note that $\lbrep$ is also a lower bound on the distillable entanglement of $\mathcal{E}_P( \Phi)$ which is achievable with a distillation protocol given in Ref.~\cite{VW03}.

Since a logical qudit is encoded into $n$ physical qudits, the number of photonic modes used to connect two repeater stations is $M=n$ and $M=nD$ for Fock and MM encoding, respectively.
Thus, the multimode PLOB-repeaterless upper bound, recall Eq.~\eqref{eq:repeaterless_PLOB_bound}, is given by
\begin{align}\label{eq:ubPLOB}
 \ubPLOB := \begin{cases}
	     - n \times \log_2(1-\eta), & \text{Fock encoding} \\
             - nD\times \log_2(1-\eta), & \text{MM encoding} \hspace{1em},
            \end{cases}
\end{align}
where $\eta$ is the transmissivity of a pure-loss channel corresponding to the total distance $L$ from Alice to Bob.
Every direct transmission channel employing the same number of photonic modes, as the quantum repeater has a quantum capacity smaller than $\ubPLOB$. 
The quantum repeater, on the other hand, has a quantum capacity larger than $\lbrep$.
Hence, the quantum repeater is genuine if (but not necessarily only if) the \emph{quantum-repeater gain},
\begin{align}\label{eq:QRgain}
 \Delta:=\lbrep - \ubPLOB,
\end{align}
is positive.

%% file: 3Results.tex
As we have argued in Sec.~\ref{sec:repeater_protocol}, error-corrected, one-way qudit repeaters have a quantum capacity of at least $\lbrep = \log_2(D)-H(P)$, where the error probability distribution  $P=(p^{\mathrm{fin}}_{r,s})_{r,s\in\ZDZ}$ depends on various parameters of the quantum repeater such as
the qudit dimension $D$, the distance $L$ between Alice and Bob, the number $N-1$ of repeater stations, and the parameters of the $\llbracket n,1,d \rrbracket_D$  QECC. 
In Sec.~\ref{sec:error_model}, we introduce our noise model and derive a worst-case approximation of the pure-loss channel for Fock qudits. 
In Sec.~\ref{sec:stats}, we provide the analytical dependence of $P$ on all these parameters for both Fock and MM encoding.
Afterwards, in Secs.~\ref{sec:optimizing_delta}$-$\ref{sec:resource_estimation}, we investigate parameter regimes for genuine quantum repeaters.

\subsection{Noise model} 
\label{sec:error_model}

In a realistic scenario, Alice and Bob have to deal with errors. 
Each transmission channel from one repeater station to the next is modeled as a pure-loss channel $\mathcal{E}^{(\eta_0)}_\mathrm{loss}$ with transmissivity
\begin{align}\label{eq:transmission_errors}
\eta_0 = 10^{-\frac{\alpha L_0}{10}},
\end{align}
where $L_0$ is the distance between the repeater stations and $\alpha = 0.2$ dB/km is the attenuation of optical fibers at 1550 nm  wavelength~\cite{GRTZ02}. 
We derive an approximation of pure-loss channels with generalized Pauli-error channels in Sec.~\ref{sec:pauli_approx} as preparation for an error analysis using the framework of Ref.~\cite{MHKB18}.

Besides transmission losses, we also include measurement errors ($f_\mathrm{M}$), modeled by a depolarizing channel $\mathcal{E}^{(1-f_\mathrm{M})}_{\mathrm{depol};D}$, recall Eq.~\eqref{eq:depol}, before each measurement. 
Unless stated otherwise, we use the value $f_\mathrm{M}=0.01$, as the best single-photon detector efficiencies of about $95\%$ match this order of magnitude~\cite{Hadfield09}.
To model gate errors ($f_\mathrm{G}$), we furthermore assume that all $\CZ$ gates are followed by two depolarizing channels (one on each qudit) with an optimistic error parameter $f_\mathrm{G}=10^{-3}$. 
Deterministic photon-photon gates for polarization qubits based on light-matter interactions have been demonstrated with an average gate fidelity of $76.2\pm 3.6\%$~\cite{HWRR16}.
Two-mode gates for Fock qudits could in principle be realized using Kerr-interactions~\cite{ADGJ01}, however, 
high-fidelity phase gates have only been reported for a single photonic mode~\cite{HVHKAFLS15}.
In Sec.~\ref{sec:device_errors}, we will examine how the quantum-repeater gain $\Delta$ depends on the operational error rates $f_\mathrm{M}$ and $f_\mathrm{G}$.

Finally, storage errors ($f_\mathrm{S}$) affecting Alice's physical qudits in the QMs are modeled by depolarizing channels, $\mathcal{E}^{(1-f_\mathrm{S})}_{\mathrm{depol};D}$,
with
\begin{align} \label{eq:storage_errors}
 1- f_S = 10 ^{- \frac{\gamma L/c}{10}},
\end{align}
where $\gamma$ is the decaying rate of Alice's QM and $c= 200\mathrm{km/ms} $ is the speed of light in a fiber with  a refractive index of 1.5.
Optical fiber loop QMs, with $\gamma_\mathrm{fiber}= \alpha c =40\mathrm{dB/ms}$, are not useful, as the stored qudits decay with the same rate as the flying qudits. 
However, matter-based QMs have been demonstrated:
Cold atomic ensembles provide QMs with $\gamma_\mathrm{atom}=5\mathrm{dB/ms}$ ($50\%$ efficiency in $0.6\mathrm{ms}$)~\cite{CCEBHCGRLB16}.  
Promising candidates are based on nitrogen vacancy centers in diamond which range from 
$\gamma_\mathrm{NV}=4\times 10^{-3}\mathrm{dB/ms}$ (coherence time $T_2\approx1\mathrm{s}$)\cite{MKLJYBPHCMTCL12, ACBKMTT18} to $\gamma_\mathrm{NV}=7\times 10^{-5}\mathrm{dB/ms}$ ($T_2\approx60\mathrm{s}$)\cite{RABDBMTT19}.
Using trapped ions, decaying rates of  $\gamma_\mathrm{ion} = 7\times10^{-6} \mathrm{dB/ms}$ are possible
\cite{WUZALZDYK17}.
Since these proof-of-principle QMs do not take storage-and-retrieval efficiencies into account,  we use a more realistic value of $\gamma = 10^{-2} \mathrm{dB/ms}$ ($T_2\approx100\mathrm{ms}$~\cite{KMLNRR18}) for our analysis.

\subsubsection{Approximation of pure-loss channels with generalized Pauli-channels}
\label{sec:pauli_approx}

{For our error analysis it is crucial that all error channels are modeled as Pauli error channels.}
Recall from Eq.~\eqref{eq:erase} that the pure-loss channel, $\mathcal{E}_\mathrm{loss}^{(\eta_0)}$, acts on MM qudits as an erasure channel. 
{If we do not use the information of an occurred erasure, i.e., we replace the flag $\ket{\mathrm{vac}}\bra{\mathrm{vac}}$ by the completely mixed state $\mathbbm{1}/D$, we can replace  $\mathcal{E}_\mathrm{loss}^{(\eta_0)}$ by $\mathcal{E}_{\mathrm{depol};D}^{(\eta_0)}$.
Note that this simplification cannot overestimate the performance of the quantum repeater.
}

If, on the other hand, the qudits are encoded in the Fock basis, the pure-loss channel introduces errors on the number of photons.
Losing exactly $r$ photons can be regarded as an application of the error operator $E=X^{-r}$, recall Eq.~\eqref{eq:PauliXZ}.
For a given input state $\ket{k}$, the probability for this to happen is given by
\begin{align}\label{eq:prob_pure_loss_Fock}
\mathrm{Pr} \left( E=X^{-r} \ \big \vert \ \rho_\mathrm{in} = \ket{k}\bra{k}\right) 
 = \begin{cases} 
   {k \choose r} \eta_0^{k-r} (1-\eta_0)^r, & r\le k \\
  0, & r>k
  \end{cases},
\end{align}
where $k,r\in\{0,\ldots,D-1\}$, as we have seen in Eq.~\eqref{eq:lossy_on_fock}.
To upper bound the $X$ error probabilities, we set
\begin{align}\label{eq:Pauli_approx_Fock_coeff}
p^\mathrm{appr}_{-r}:=  \max_{k\in\{0,\ldots,D-1\}} \mathrm{Pr}\left (E=X^{-r} \ \big \vert \ \rho_\mathrm{in}=\ket{k}\bra{k} \right) 
\end{align}
for all $r\neq 0$.
If  $p^\mathrm{appr}_{0} := 1- \sum_{r=1}^{D-1} p^\mathrm{appr}_{-r}$ is positive, 
we can model the pure-loss channel on Fock state qudits of a single bosonic mode by the generalized Pauli-error channel,
\begin{align}\label{eq:Pauli_approx_Fock}
 \mathcal{E}^{(\eta_0)}_{\mathrm{appr};D} : \rho \longmapsto \sum_{r=0}^{D-1} p^\mathrm{appr}_{-r} X^{-r} \rho (X^{-r})^\dagger ,
\end{align} 
as a worst-case approximation.
In the error analysis for Fock-state encoding, we thus replace each pure-loss channel,$\mathcal{E}_\mathrm{loss}^{(\eta_0)}$, between any two repeater stations by $\mathcal{E}^{(\eta_0)}_{\mathrm{appr};D}$. 
Note that this further decreases $\lbrep$; thus, $\Delta>0$ will still indicate a genuine quantum repeater.
We will only consider repeater lines for which the repeater stations are spaced close enough such that $p^\mathrm{appr}_0\ge0$, i.e., $\eta_0 \approx 1$.
To compute the error probabilities in Eq.~\eqref{eq:Pauli_approx_Fock_coeff}, we have to find the input state $\ket{k}$ with the highest probability to lose $r$ photons.
By differentiating the analytical continuation of Eq.~\eqref{eq:prob_pure_loss_Fock}, we obtain
\begin{align}\label{eq:critical_k}
0&= \frac{\partial}{\partial k} \left[{k \choose r} \eta_0^{k-r} (1-\eta_0)^r \right] 
 = {k \choose r} \eta_0^{k-r} (1-\eta_0)^r \left( \ln(\eta_0) + H_k - H_{k-r} \right ) ,
\end{align}
where
\begin{align}
 H_k  := \sum_{j=1}^k \frac{1}{j}=\ln(k) + \gamma_\mathrm{EM} +\frac{1}{2k} -\frac{1}{12k^2} +\frac{1}{120k^4}-\ldots
\end{align}
is the $k$th Harmonic number and  $\gamma_\text{EM}\approx 0.577$ is the Euler-Mascheroni constant.
For large $k$, the approximation $H_k - H_{k-r} \approx \ln(k)-\ln(k-r)$ yields  $k\approx {r}/({1-\eta_0})$ as the solution of Eq.~\eqref{eq:critical_k}. 
This is indeed a maximum because the sign of the derivative given in Eq.~\eqref{eq:critical_k} changes at $k\approx r/(1-\eta_0)$ from plus to minus, when increasing $k$.
This follows from the positivity and the strictly monotonic decrease of the derivative of the analytical continuation of $H_k$.
Let $\round(x)$ denote the integer that is closest to $x\in\mathbb{R}$.
Because of $\eta_0 \approx 1$, the approximation is so good 
that the integer
$\tilde k(r,\eta_0):= \min\{\round\left(r/({1-\eta_0})\right), D-1\}$  
is the number of input photons having the highest probability (among inputs of up to $D-1$ photons) to lose exactly $r$ photons. 
This yields 
\begin{align}\label{eq:pappr}
 p^\mathrm{appr}_{-r} =  {\tilde k(r,\eta_0) \choose r} \eta_0^{\tilde k(r,\eta_0)-r}(1-\eta_0)^{r}
\end{align} 
as the solution of Eq.~\eqref{eq:Pauli_approx_Fock_coeff}.

\subsection{Error statistics for error-corrected, one-way qudit repeaters}
\label{sec:stats}
In order to evaluate the quantum-repeater gain defined in Eq.~\eqref{eq:QRgain}, one has to know the error probability distribution $P$ of the entangled state distributed by the quantum repeater. 
In Sec.~\ref{sec:stats_multimode} and Sec.~\ref{sec:stats_fock}, we derive for MM and Fock qudits, respectively, the expression of the success probabilities $p_\mathrm{succ}^\mathrm{1.rep}$, $p_\mathrm{succ}^\mathrm{rep}$, $p_\mathrm{succ}^{\mathrm{Bob},X}$, and $p_\mathrm{succ}^{\mathrm{Bob},Z}$ (recall the paragraph above Eq.~\eqref{eq:p_fin_x} for their definitions) from which $P$ follows via  Eqs.~\eqref{eq:p_fin_x} and \eqref{eq:p_fin_z}.

\subsubsection{Error statistics for multimode qudits}
\label{sec:stats_multimode}
 
 
Here, we review our previous results~\cite{MHKB18} which hold for error-corrected qudit repeater lines where the qudits are encoded into multiple bosonic modes, e.g., time-bin, TM, and OAM qudits.
{Recall that we replace the pure-loss channel with transmissivity $\eta_0$ by a  depolarizing channel with transmission error rate $f_\mathrm{T} = 1-\eta_0$.}   
Depolarizing channels can be regarded as sources of discrete $X$ and $Z$ errors which propagate through the repeater line.
It turns out that there are six sources from which a $Z^{e_i}$ error at the $X$ measurement of qudit $i$ can originate: two transmission, three gate, and one measurement error channel~\cite{EKB16_analysis, EKB16_networks, EKB16_router, MHKB18}.
Such an error will change the physical measurement result $c_i$ into $c_i+e_i\in\ZDZ$ with probability $p^\mathrm{rep}_{e_i}$, where 
\begin{align} \label{eq:p_rep_MM} 
p^\mathrm{rep}_{{e_i}\neq 0} &= \frac{1}{D} \left[ 1- (1-f_\mathrm{T})^2 (1-f_\mathrm{G})^3 (1-f_\mathrm{M}) \right ]
\end{align}
and $p^\mathrm{rep}_0 = 1-(D-1)p^\mathrm{rep}_{e_i\neq 0}$.
For the first repeater station, fewer error sources contribute: 
$p^{1.\mathrm{rep}}_{e_i\neq 0} = \frac{1}{D} \left[ 1- (1-f_\mathrm{T}) (1-f_\mathrm{G})^2 (1-f_\mathrm{M}) \right ]$ and $p^\mathrm{1. rep}_0 = 1-(D-1)p^\mathrm{1. rep}_{e_i\neq 0}$.
Similarly, the probability of an $X^{e_i}$  and $Z^{e_i}$ error on Bob's qudit right before the stabilizer measurement is given by $p^{\mathrm{Bob},X}_{e_i}$ and $p^{\mathrm{Bob},Z}_{e_i}$, respectively, where
$p^{\mathrm{Bob},X}_{e_i\neq0} =\frac{1}{D}\left[1-(1-f_\mathrm{G})^2(1-f_\mathrm{S})\right],$ and 
$p^{\mathrm{Bob},Z}_{e_i\neq0} =\frac{1}{D}\left[1-(1-f_\mathrm{T})(1-f_\mathrm{G})^3(1-f_\mathrm{S})\right]  , $
and $p^{\mathrm{Bob},X}_0$, $p^{\mathrm{Bob},Z}_0$ again follow from normalization.
See Ref.~\cite{MHKB18} for more details.  

In the following, let $p_{e_i}$ be either $p^{\mathrm{rep}}_{e_i}$, $p^{1.\mathrm{rep}}_{e_i}$,  $p^{\mathrm{Bob},X}_{e_i}$, or $p^{\mathrm{Bob},Z}_{e_i}$, and likewise for $p_\mathrm{succ}$.
In either situation, $n$ individual measurement results are employed for the error correction attempt based on the $\llbracket n,1,d\rrbracket_D$ QECC.
Since the error probability of a single error $e_i\in\ZDZ$ is given by $p_{e_i}$, the probability of an error pattern $\textbf{e}=(e_1,\ldots,e_n)$  at the respective error correction attempt is given by 
\begin{align} \label{eq:p_pattern}
 p_\textbf{e} = \prod_{i=0}^np_{e_i} = p_0^{n-\wt(\textbf{e})} p_{e\neq0}^{\wt(\textbf{e})},
\end{align}
where the Hamming weight $\wt(\textbf{e})$ denotes the number of nonzero entries of $\textbf{e}$.
As the QECC can correct up to $\lfloor (d-1)/2\rfloor$ errors, the probability of a correctable error pattern is given by
\begin{align}\label{eq:p_succ_MM}
 p_\mathrm{succ} &= \sum_{k=0}^{\left \lfloor\frac{d-1}{2}\right \rfloor} (D-1)^k{n \choose k} p_0^{n-k} p_{e\neq0}^k.
\end{align}
If an error pattern occurs which cannot be corrected, a logical error is guessed with 
probability $p_\mathrm{guess}={(1-p_\mathrm{succ})/D}$.
Combining the respective success probabilities according to Eqs.~\eqref{eq:p_fin_x} and \eqref{eq:p_fin_z} yields the final error distribution on the distributed state.

\subsubsection{Error statistics for Fock qudits}
\label{sec:stats_fock}

Here, we adapt the error analysis of Ref.~\cite{MHKB18} to error-corrected qudit repeater lines with physical qudits encoded in the Fock basis of a single bosonic mode.
In this case, the propagation of errors is more complicated, as the error probabilities of  $\mathcal{E}^{(\eta_0)}_{\mathrm{appr};D}$ all differ from each other.
A logical $\CZ$ gate for the $\llbracket n,1,d\rrbracket_D$ quantum polynomial code is transversal in the sense that there are invertible elements $s_1,\ldots,s_n \in \ZDZ$ such that 
$\bigotimes_{i=1}^n \CZ ^{s_i}(a_i,b_i)$ acts as a $\CZ$ gate between two logical qudits $a$ and $b$,
where $\CZ(a_i,b_i)$ denotes a physical $\CZ$ gate from the $i$th physical qudit of the logical qudit $a$ to the $i$th qudit of $b$. 
For MM qudits, in the previous section, this is not important because, at a depolarizing channel, every nontrivial error occurs with the same probability. 
Here, however, an $X^{s_i^{-1}e_i}$ error, occurring during the transmission to physical qudit $i$, will induce a $Z^{e_i}$ error to qudit $i$ of the next logical qudit for all $e_i\in\ZDZ$.
Employing the error tracking tools of Ref.~\cite{MHKB18} and taking all relevant error sources into account,
we obtain that the probability for an error $e_i$ on the measurement result of qudit $i$ at every repeater station but the first is given by
\begin{align} \label{eq:p_rep_Fock}
 p^\mathrm{rep}_{e_i} &= p^\mathrm{appr}_{s_i^{-1}e_i} (1-f_\mathrm{G})^3(1-f_\mathrm{M})  + \frac{1}{D} \left[ 1- (1-f_\mathrm{G})^3 (1-f_\mathrm{M}) \right],
\end{align}
recall Eq.~\eqref{eq:pappr} for the definition of  $p^\mathrm{appr}_{-r}$.
For the first repeater station,  
\begin{align} 
 p^{1.\mathrm{rep}}_{e_i\neq 0} &= \frac{1}{D} \left[ 1- (1-f_\mathrm{G})^2 (1-f_\mathrm{M}) \right]
\end{align}
and $p^{1.\mathrm{rep}}_0$ follows from normalization; note that transmission errors do not contribute, as they are of $X$ type for the channel $\mathcal{E}^{(\eta_0)}_{\mathrm{appr};D}$.
Similarly, $p^{\mathrm{Bob},X}_{e_i}$ and $p^{\mathrm{Bob},Z}_{e_i}$  are given by
\begin{align} 
p^{\mathrm{Bob},X}_{e_i\neq0} &=\frac{1}{D}\left[1-(1-f_\mathrm{G})^2(1-f_\mathrm{S})\right], 
\\ \text{and}  \hspace{1em}
 p^{\mathrm{Bob},Z}_{e_i} &= p^\mathrm{appr}_{s_i^{-1}e_i}(1-f_\mathrm{G})^3(1-f_\mathrm{S}) +\frac{1}{D} \left[ 1-(1-f_\mathrm{G})^{3}(1-f_\mathrm{S})  \right].	 \nonumber
\end{align}

This time, let $p_{e_i}$ be either  $p^{\mathrm{rep}}_{e_i}$ or $p^{\mathrm{Bob},Z}_{e_i}$  
(for $p^{1.\mathrm{rep}}_{e_i}$ and   $p^{\mathrm{Bob},X}_{e_i}$ we can continue as in Sec.~\ref{sec:stats_multimode}), 
and likewise for $p_\mathrm{succ}$.
Again, the probability of an error pattern $ \textbf{e}= (e_1,\ldots, e_n)$ is given by $p_\textbf{e}=\prod_{i=1}^n p_{e_i}$, but here we cannot simplify this expression using the Hamming weight because the nontrivial error probabilities do not coincide.
The probability that a correctable error pattern occurs is given by
the sum over all probabilities $p_\mathbf{e}$ where $\wt(\mathbf{e})\le \lfloor (d-1)/2\rfloor$.
Since the substitution $e_i' := s_i^{-1}e_i$ does not change the Hamming weight, this sum does not depend on the $s_i$ and can be expressed as
\begin{align}\label{eq:p_succ_fock}
 p_\mathrm{succ} &= \sum_{k=0}^{\left \lfloor \frac{d-1}{2}\right \rfloor} {n \choose k} p_0^{n-k} \left( \sum_{\textbf{r}\in\{1,\ldots,D-1\}^k}   p_\textbf{r} \right ),
\end{align}
where  $p_\textbf{r} := p_{r_1}p_{r_2}\ldots p_{r_{k}}$. 
By combining terms with equal probability in the inner sum over {all} combinations of nontrivial error patterns $\textbf{r}=(r_1,\ldots,r_k)$, we find 
\begin{align}\label{eq:combi_sum}
\sum_{\textbf{r}\in\{1,\ldots,D-1\}^k}   p_\textbf{r} &=  \sum _{\ell_1+\ldots+\ell_{D-1}=k} {k \choose \ell_1, \ldots, \ell_{D-1}} p_{1}^{\ell_1}  \ldots p_{D-1}^{\ell_{D-1}}  ,
\end{align}
where for $\ell_1+\ldots+\ell_{D-1}=k$ the multinomial coefficient is defined as 
\begin{align}
 {k \choose \ell_1, \ldots, \ell_{D-1}} &= \frac{k!}{\ell_1!\ldots \ell_{D-1}!}.
\end{align} 
Note that $s_1=\ldots=s_n=1$ can be assumed for the evaluation of Eq.~\eqref{eq:combi_sum}.
Because $p_{e_i}\neq p_{{e'}_i}$  for $e_i \neq {e'}_i$,  no further simplification can be made through combining coinciding terms.
As before, the final error distribution follows from the corresponding success probabilities.

\subsection{Optimizing the quantum-repeater gain} 
 \label{sec:optimizing_delta}
  
\begin{figure}[t]
  \centering
  \begin{minipage}{4em}
   \flushright $\Delta$
   \end{minipage}
  \begin{minipage}{26em}
  \includegraphics[width = \textwidth]{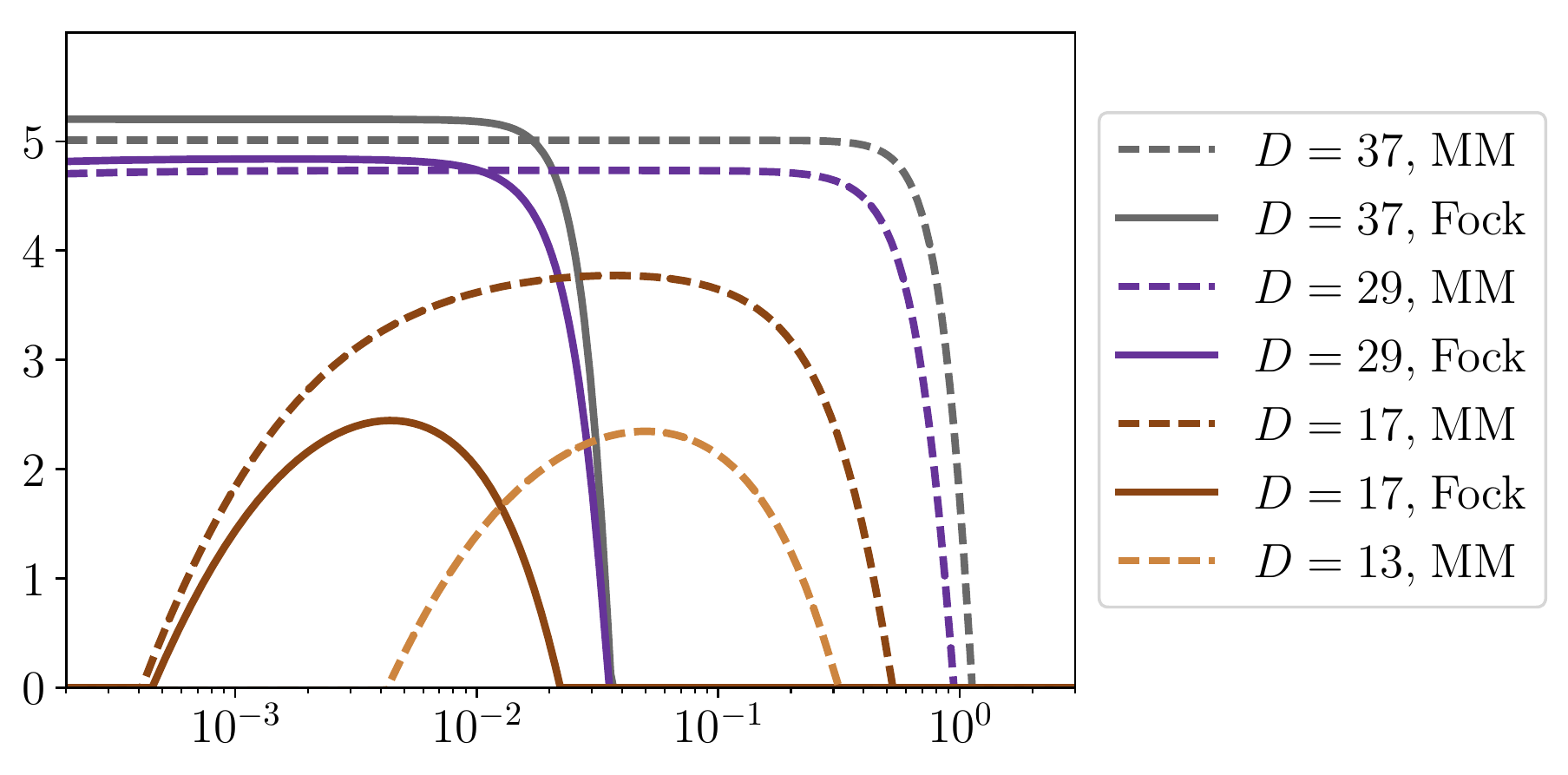}
  \end{minipage} 
   
   \hspace{-2.5em}
   $L_0$ in km
  \caption{Quantum-repeater gain $\Delta=\lbrep-\ubPLOB$  in terms of the spacing $L_0$ between adjacent repeater stations for a quantum repeater line of total length $L=200$km. 
  The qudits are encoded with a $\llbracket D,1,(D+1)/2\rrbracket_D$ QECC where the qudit dimension $D$ is color coded. 
  Dashed and solid curves correspond to MM and Fock qudits, respectively.
  Note that $\Delta$ is negative for $D=13$ Fock qudits because it cannot be ensured that sufficiently many errors can be corrected.
  We use the error model of Sec.~\ref{sec:error_model} with $\alpha =0.2\mathrm{dB/km}$, $f_\mathrm{M}=10^{-2}$, $f_\mathrm{G}=10^{-3}$, and $\gamma = 10^{-2}\mathrm{dB/ms}$, i.e., $f_\mathrm{S}\approx 2.3\times10^{-3}$.
}
  \label{fig:Delta_vs_L0}
\end{figure}

In order to identify genuine quantum repeaters, we want to find parameter regions where the quantum-repeater gain, $\Delta=\lbrep-\ubPLOB$, takes values which are significantly larger than zero. 
The first parameter we focus on is the repeater spacing $L_0$.
In Fig.~\ref{fig:Delta_vs_L0}, 
$\Delta$ is plotted as a function of $L_0$
for a quantum repeater line of total length $L=200\mathrm{km}$,
a distance large enough that $\Delta$ takes positive values while the $M$-mode PLOB-repeaterless bound,
$\ubPLOB \approx 1.44M \times 10^{-4}$, still has a recognizable influence.
The selected QECCs have parameters $\llbracket D,1,(D+1)/2\rrbracket_D$, i.e., they saturate the quantum singleton bound. Note that this also is the largest possible code distance for which quantum polynomial codes are available at a given dimension $D$, where $D$ is a prime number.
Thus, the considered QECCs are the best available for a given qudit dimension.
The prime dimensions $D\in\{13,17,29,37\}$ are selected such that the code distance $d=(D+1)/2$ of the QECC is odd, as this ensures that the number of correctable single qudit errors is $t=(d-1)/2$ (and not $t=(d-2)/2$).
For large $D$, the error correction capabilities perform sufficiently good, such that $H(P)$, the Shannon entropy of the error probability distribution, is almost zero, i.e., Alice and Bob have almost perfect knowledge about the state of their qudits.
In this saturated regime,  where $\Delta$ does not significantly change with respect to $L_0$ over some  orders of magnitude, 
the height of the plateau, $\max_{L_0}\Delta = \log_2(D) - H(P) -\ubPLOB  \approx  \log_2(D)  -\ubPLOB $, is, by Eq.~\eqref{eq:ubPLOB}, larger for Fock qudits than for MM qudits because logical Fock qudits employ only $M=D$ modes while MM qudits require $M=D^2$ modes.
Thus, for short distances $L_0$, the repeaterless quantum capacity is larger for MM qudits.
Indeed, the gap between the Fock and the MM plateau is given by
\begin{align}
\ubPLOB(\mathrm{MM}) -\ubPLOB(\mathrm{Fock})
\approx  
1.44\times 10^{-4}\times
(D^2-D) &\approx \begin{cases}
                           0.2 & \text{ for } D=37\\
                           0.1 & \text{ for } D=29 
                           \ \ .
                          \end{cases}
\end{align}
As $L_0$ further increases, transmission losses start to significantly deteriorate the error correction procedure, causing a sudden drop of $\Delta$. 
The largest possible repeater spacing for which the PLOB-repeaterless bound is surpassed is on the order of $L_0\sim1$km for MM qudits and 
$L_0\sim0.01-0.1$km
for Fock qudits, respectively.
For Fock qudits, the quantum-repeater gain is more vulnerable to transmission losses because of our worst-case approximation of the corresponding pure-loss channel, recall Eq.~\eqref{eq:Pauli_approx_Fock}.
On the other hand, for repeater lines with a very small repeater spacing $L_0$,
operational errors (gate and measurement errors) start to play a critical role, as 
more repeater stations increase the number of error sources, until eventually
the lower bound on the repeater's quantum capacity, $\lbrep$, vanishes and $\Delta$ coincides with $-\ubPLOB$. 
Since we assume depolarizing noise for operational errors in both encodings, repeaters based on MM and Fock qudits qualitatively show the same behavior for small $L_0$.
For small $D$, the code distance $d$ 
is too small, both transmission losses and operational errors deteriorate the error correction procedure, which prohibits the formation of a plateau where $\Delta(L_0)$ is constant. 
We stress that the repeater spacing $L_0$ can be raised tremendously if the intended quantum-repeater gain $\Delta$ is sub-optimal, e.g., $\lbrep = 0.9 \times B^{\downarrow \mathrm{max}}_\mathrm{rep}$.

\begin{figure}[t]
  \centering  
   
   \begin{minipage}{\textwidth}
   \hspace{6em} 
   Multimode qudits
   \hspace{10.5em}
   Fock qudits
   \end{minipage}

  \begin{minipage}{.02\textwidth}
   \flushright $D$
   \end{minipage} 
  \begin{minipage}{.42\textwidth}
  \includegraphics[height= 13.5em]{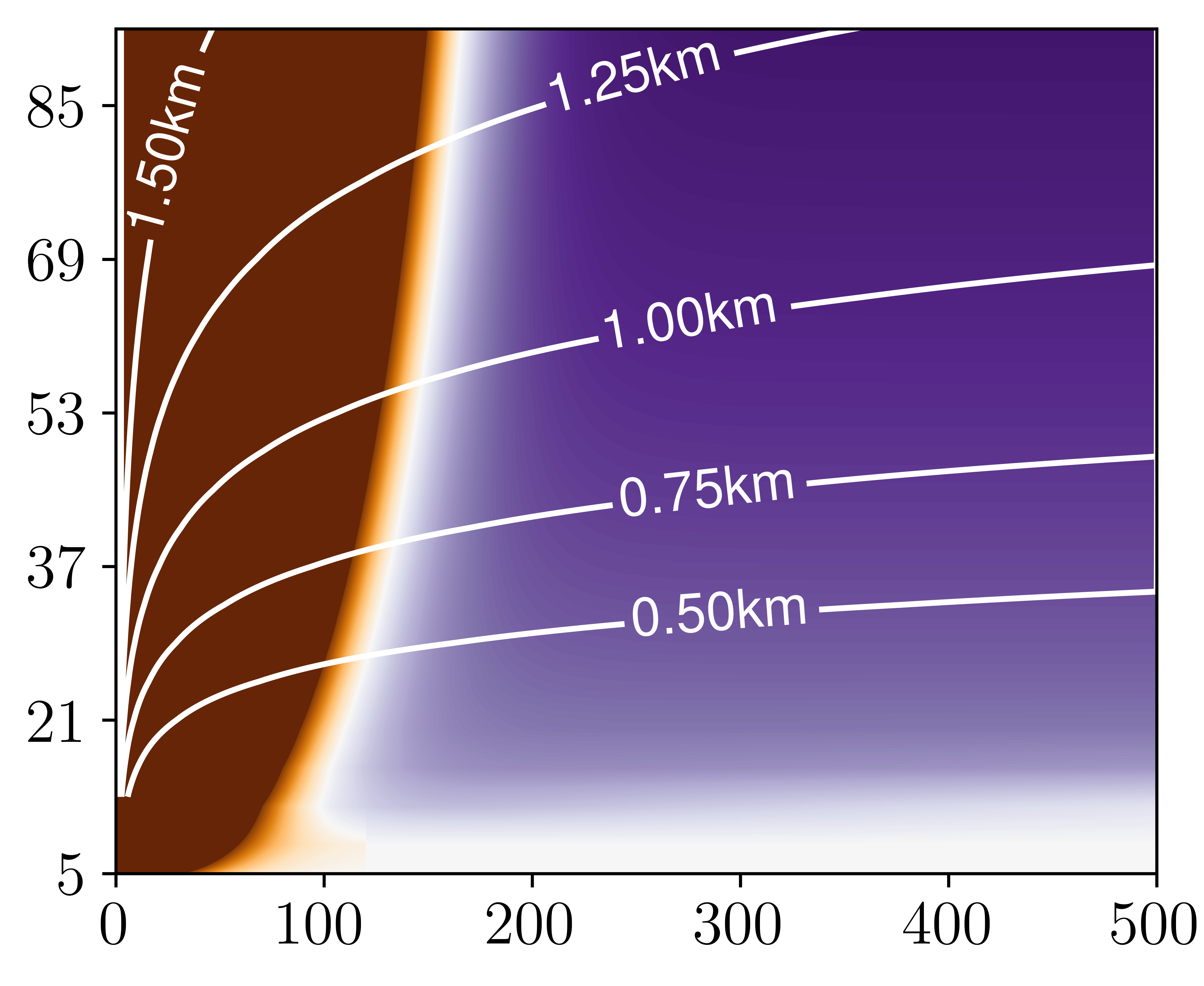} 
  \end{minipage}
  \begin{minipage}{.015\textwidth}
   \flushright $D$
   \end{minipage} 
  \begin{minipage}{.46\textwidth}
  \includegraphics[height= 13.5em]{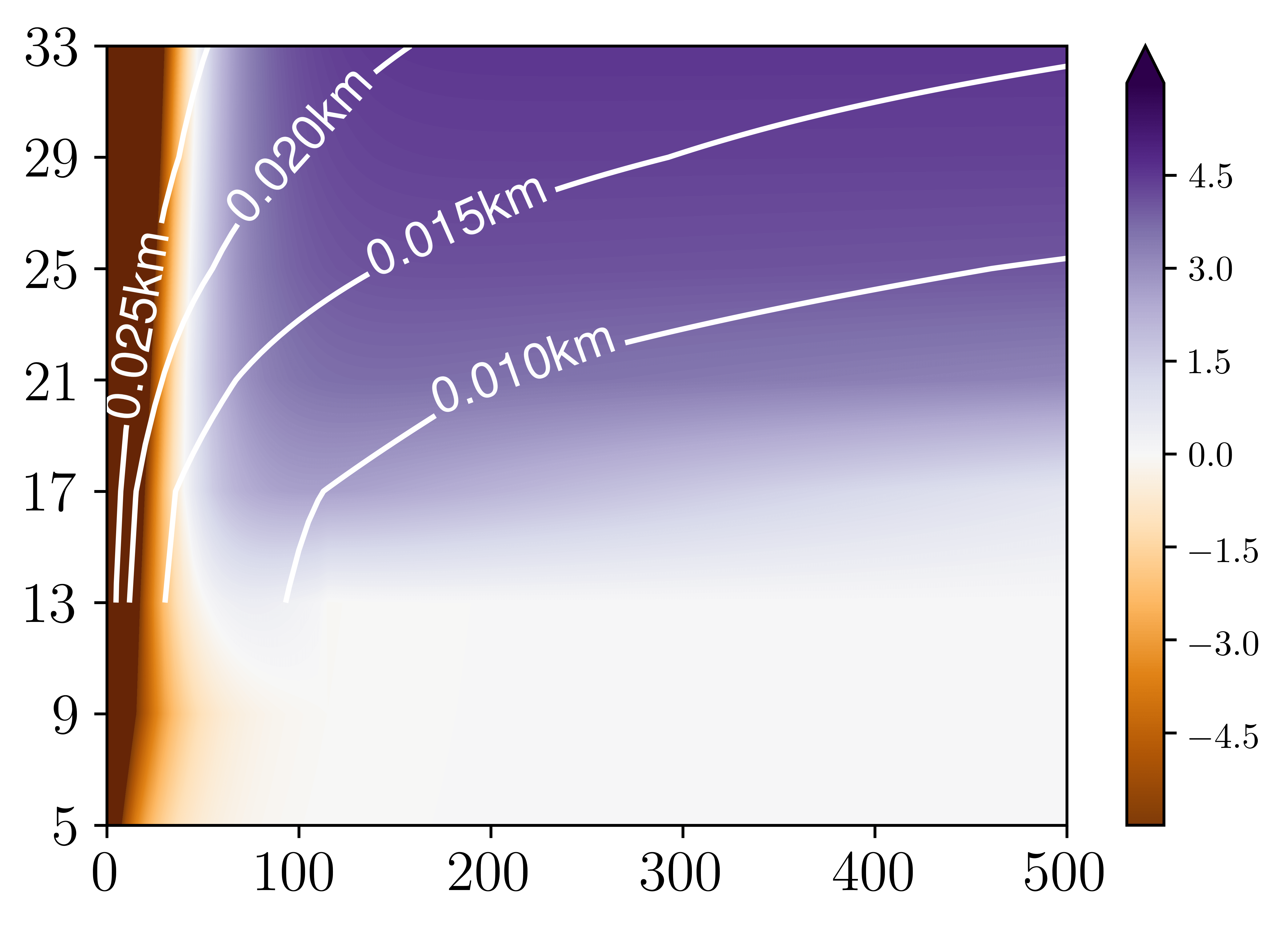} 
   \end{minipage} 
  \begin{minipage}{.02\textwidth}
   $\Delta$
   \end{minipage}

   \begin{minipage}{\textwidth}
   \hspace{8em} 
   $L$ in km 
   \hspace{13.2em}
   $L$ in km 
   \end{minipage}
 
  \caption{Quantum-repeater gain $\Delta=\lbrep-\ubPLOB$  and corresponding repeater spacing $L_0$ (white lines) for $D$-dimensional qudits based on MM (left) and Fock (right) encoding, 
  The qudits are encoded with a  $\llbracket D,1,(D+1)/2\rrbracket_D$ QECC where the qudit dimension varies in steps of 4  from $D=5$ to $D=93$ and $D= 33$ for MM and Fock encoding, respectively.  
Note that the computational complexity of Eq.~\eqref{eq:p_succ_fock} limits $D\le33$ for Fock qudits.
The total distance between Alice and Bob varies from $L=0\mathrm{km}$ to $L=500\mathrm{km}$ and the repeater spacing $L_0= L/N$ is adjusted such that $\lbrep = 0.9\times B^{\downarrow \mathrm{max}}_\mathrm{rep}$.
  We use the error model of Sec.~\ref{sec:error_model} with $\alpha =0.2\mathrm{dB/km}$, $f_\mathrm{M}=10^{-2}$, $f_\mathrm{G}=10^{-3}$, and $\gamma = 10^{-2}\mathrm{dB/ms}$.
}
  \label{fig:Delta_vs_L_and_D}
\end{figure}

In Fig.~\ref{fig:Delta_vs_L_and_D}, 
we display the quantum-repeater gain $\Delta$ (color coded) for quantum repeater lines of varying total length $L$ (abscissa) and qudit dimension $D$ (ordinate). 
We vary  $D$ in steps of 4 such that the considered  $\llbracket D,1,(D+1)/2\rrbracket_D$ QECCs have an odd code distance $d$, in between, we interpolate.
The corresponding values of $L_0$ are included in Fig.~\ref{fig:Delta_vs_L_and_D} via white contour lines.
For MM and Fock qudits, respectively, the repeater spacing is on the order of $L_0\sim 1\mathrm{km}$ and  $L_0\sim 0.01\mathrm{km}$, respectively. 
If the total length $L$ of the repeater line is shortened, transmission losses become less important and operational errors begin to dominate. 
Thus, to reach $\lbrep= 0.9\times B^{\downarrow \mathrm{max}}_\mathrm{rep}$,
the spacing $L_0$ between two adjacent repeater stations is increased, as this decreases the number of operational error sources.
The spacing $L_0$ also increases with $D$ because QECCs with a higher code distance $d= (D+1)/2$ can correct more errors. 

We find three distinct regions in Fig.~\ref{fig:Delta_vs_L_and_D}, each with a typical signature:
(i) For small $L$, the PLOB-repeaterless bound cannot be surpassed, i.e., $\Delta<0$.
(ii) For large $L$ and a large code distance $d=(D+1)/2$ we observe $\Delta>0$.
(iii) For large $L$ and small $d$ we find $\Delta\approx 0$.
We now discuss the signatures of these three regions: 
\begin{itemize}
\item[(i)] [$\Delta <0$] At the brown region on the left, the $M$-mode PLOB-repeaterless bound is much larger than $\log_2(D)\ge\lbrep$.
%
Asymptotically, it is even unbounded,
\begin{align}
 \ubPLOB =-M\times \log_2(1-\eta) \overset{L\rightarrow 0}{\longrightarrow} \infty.
\end{align}
As logical $\llbracket D, 1, (D+1)/2 \rrbracket_D$ MM qudits are encoded into $M=D^2$ modes, this region extends to longer total lengths $L$ of the repeater line if the qudit dimension $D$ is larger, e.g., 
$L(\Delta<0,D=13)\lesssim100\mathrm{km}$ and $L(\Delta<0,D=85)\lesssim 150\mathrm{km}$.
For Fock qudits, which only require $M=D$ modes, this effects is barely noticeable and $L(\Delta<0)\lesssim 50\mathrm{km}$ for all $D$.

\item[(ii)] [$\Delta >0$] At the purple region on the upper right, quantum repeaters can surpass the PLOB-repeaterless bound because $\ubPLOB \approx 0$ while $\lbrep >0$.
The quantum-repeater gain $\Delta = \lbrep - \ubPLOB \approx \log_2(D)- H(P)$  increases (for a fixed $L$) in $D$ if  the distance $d=(D+1)/2$ of the QECC is large enough, as this causes $H(P)\approx 0$.
Genuine MM quantum repeaters are possible for $D\ge 13$, whereby the minimal repeater length increases with the number of modes $M=D^2$, as  already discussed for region (i).
For quantum repeaters with Fock encoding, the PLOB-repeaterless bound can be outperformed for $D\ge 17$ and $L>50\mathrm{km}$. 
For $D=13$, we observe a small quantum-repeater gain $\Delta\in(0.1, 0.5)$ for quantum repeater lines with a total length $L$ between $60\mathrm{km}$ and $110\mathrm{km}$.

\item[(iii)] [$\Delta \approx 0$] At the white region on the lower right, the total length $L$ is too large and the code distance $d$  is too small such that $\ubPLOB\approx 0$ and $\lbrep =0$, respectively.
Recall that we consider $\llbracket D,1,(D+1)/2\rrbracket_D$ quantum polynomial codes, as they have the highest code distance for a given dimension, as well as transversal $\CZ$ gates and transversal $X$ measurements.
For $D\le 9$, we find $\lbrep =0$ which implies $\Delta \approx 0$.  
\end{itemize}
Let us summarize what can be learned from Figs.~\ref{fig:Delta_vs_L0} and \ref{fig:Delta_vs_L_and_D}. 
Using higher-dimensional qudits, it is possible to reach a larger quantum-repeater gain because the ideal quantum capacity of the quantum repeater is given by $\log_2(D)$.
In many cases, this optimum  can be reached by an appropriate choice of $L_0$.
Since the code distance of the best known QECCs also grows with the qudits' dimension, a side effect of higher-dimensional qudits is the possibility to increase the distance $L_0$ between two neighboring repeater stations.
For MM qudits, $L_0$ is two orders of magnitude larger than for Fock qudits. {We attribute this} to our worst-case Pauli-approximation of the pure-loss channel for Fock qudits {which leads to a lower bound on $L_0$ for genuine quantum repeaters. We stress that this bound is not tight.} 
{Within our error model, intermediate repeater stations of a genuine quantum repeater cannot be separated by a distance exceeding $L_0 \sim 1$km. Even if all error parameters except for fiber attenuation are set to zero, $L_0$ does not improve by an order of magnitude.  
Finally, within}
our error model,
Fock and MM quantum repeaters can surpass the PLOB-repeaterless bound for $L>50\mathrm{km}$ and $L>100-150\mathrm{km}$, respectively. 
MM repeaters require a larger total length because the PLOB-repeaterless bound is higher for a larger number of modes.

\subsection{Influence of operational errors on the quantum-repeater gain} 
 \label{sec:device_errors}  
   
\begin{figure}[t]
  \centering
  \begin{minipage}{4em}
   \flushright $\Delta$
   \end{minipage}
  \begin{minipage}{26em}
  \includegraphics[width = \textwidth]{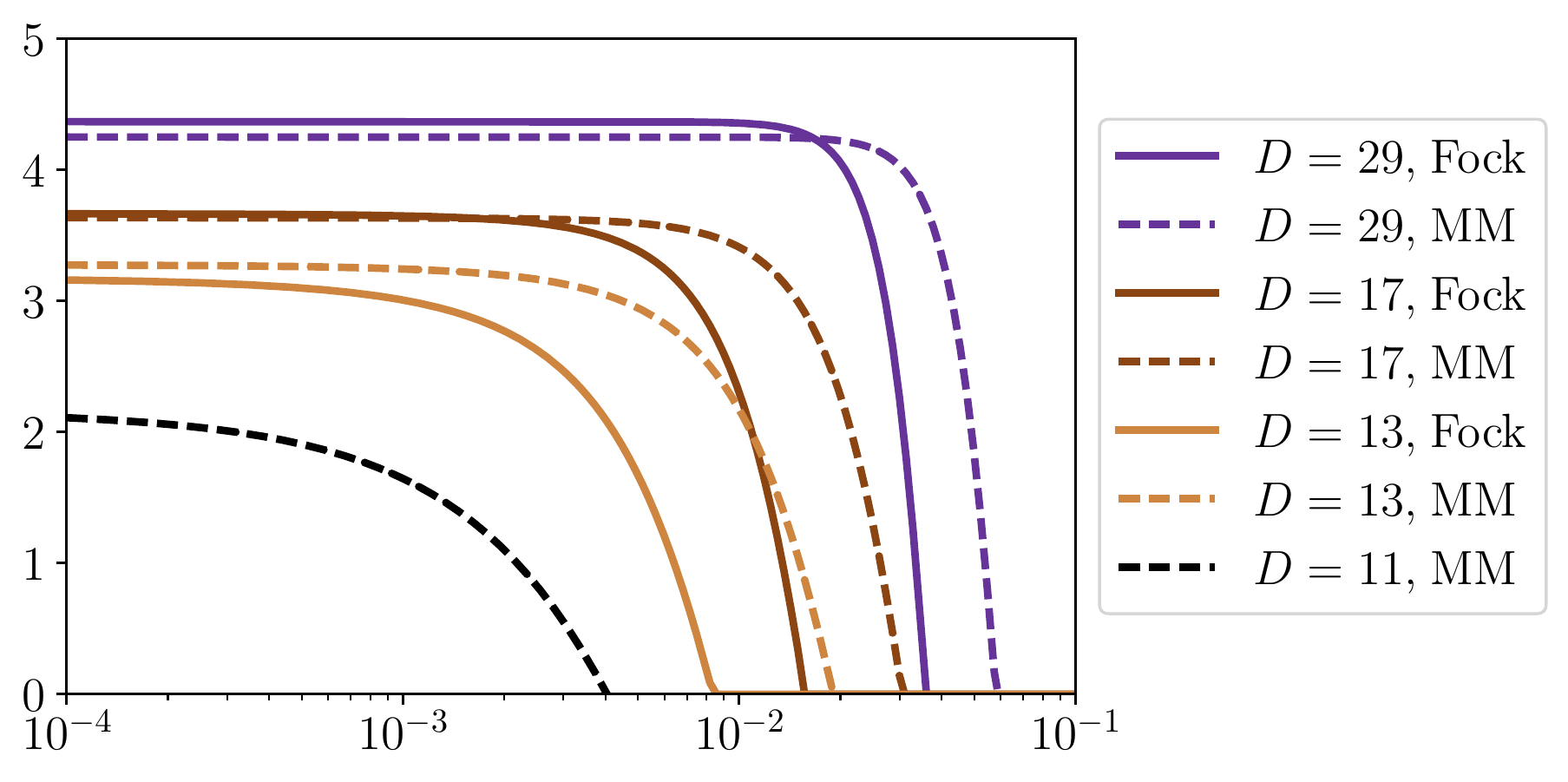}
  \end{minipage} 
   
   \hspace{-2.5em}
   $f_\mathrm{M}$ 
  \caption{The quantum-repeater gain $\Delta$ 
  in terms of the  measurement error rate $f_\mathrm{M}$.
  The other error parameters are fixed to $\alpha =0.2\mathrm{dB/km}$, $f_\mathrm{G}=10^{-3}$, and $\gamma = 0.01\mathrm{dB/ms}$, i.e., $f_\mathrm{S}\approx 2.3\times10^{-3}$.
  The qudits are encoded with a $\llbracket D,1,(D+1)/2\rrbracket_D$ QECC. 
  The repeater line has a total length of $L=200$km and the repeater spacing $L_0= L/N$ is adjusted such that $\lbrep = 0.9\times B^{\downarrow \mathrm{max}}_\mathrm{rep}$. 
}
  \label{fig:Delta_vs_fM}
\end{figure}

As it is easier to implement a $\llbracket D,1,(D+1)/2\rrbracket_D$ QECC for smaller qudit dimension $D$, 
it is desirable to lower the demands on error correction.
One way to achieve this is the improvement of operational error rates. 
As we have shown in Sec.~\ref{sec:stats}, operational errors mainly originate from intermediate repeater stations, where gate and measurement errors enter via  $(1-f_\text{G})^3$ and $(1-f_\text{M})$, respectively, recall Eqs.~\eqref{eq:p_rep_MM} and \eqref{eq:p_rep_Fock}. 
Thus, gate errors affect the quantum-repeater gain three times as large as measurement errors do, 
but they otherwise lead  to the same qualitative behavior of $\Delta$. 
Hence, we restrict the investigation of the influence of operational errors to the measurement error rate $f_\text{M}$ and fix $f_\text{G}=10^{-3}$, as before. 
Figure~\ref{fig:Delta_vs_fM}  
shows the quantum-repeater gain as a function of the measurement error rate 
$f_\text{M}$. 
We observe a similar pattern for all curves: 
At low error rates $f_\text{M}<10^{-3}$ the quantum-repeater gain $\Delta$ is almost constant. 
As $f_\text{M}$ increases, the smaller the dimension~$D$, the sooner the corresponding quantum-repeater gain drops to zero, as fewer errors can be corrected by the QECC. 
In terms of quantum-repeater gain and in direct comparison, the MM encoding is more tolerant towards measurement errors than the Fock encoded repeater line.  
As expected, lower operational error rates allow genuine quantum repeaters with smaller dimension $D$, as fewer errors need to be corrected.  
In this range of $f_\mathrm{M}$, the smallest dimension for which genuine quantum repeaters are possible is $D=11$ with MM qudits. 

\subsection{Estimate of resources} 
 \label{sec:resource_estimation}

For a resource-efficient use of quantum repeaters it is crucial to identify cost-saving candidates.
Naturally, the costs of developing and maintaining a single repeater station will increase with $D$, as a $\llbracket D,1,(D+1)/2\rrbracket_D$ QECC is employed. 
However, higher-dimensional qudits have the advantage of better error correction capabilities, thus, the number of necessary repeater stations is lower.
Table~\ref{tab:how_many_stations}
 \begin{table}[h]
 \centering 
  \begin{tabular}{lc||ccc|c}
  \hline \hline 
 & $D$  & 13 & 17 &29 & 73\\ \hline \hline 
 \textbf{MM} & $N_\mathrm{min}$ & {371} & {248} & {163} & {119} \\
  &  $L(N_\mathrm{min})$	 & 120km & 130km &140km & 170km  \\ \hline
 \textbf{Fock} & $N_\mathrm{min}$ & - & 2070 & 1705 & ? \\
 & 	 $L(N_\mathrm{min})$ &  & 56km & 60km & \\ \hline \hline 
  \end{tabular}
 \caption{The minimal number $N_\mathrm{min}$ of repeater stations for which the PLOB-repeaterless bound can be just surpassed by a quantum repeater line of a total length $L(N_\mathrm{min})$, see also Fig.~\ref{fig:how_many_stations}.
  We use the error model of Sec.~\ref{sec:error_model} with $\alpha =0.2\mathrm{dB/km}$, $f_\mathrm{M}=10^{-2}$, $f_\mathrm{G}=10^{-3}$, and $\gamma = 10^{-2}\mathrm{dB/ms}$.
 For $D=13$ Fock qudits, the PLOB-repeaterless bound is not surpassed, and for $D=73$ Fock qudits, the corresponding minimum cannot be evaluated due to the computational complexity of  Eq.~\eqref{eq:p_succ_fock}.
 }
 \label{tab:how_many_stations}
 \end{table}
provides the minimal requirement on the number of repeater stations of a genuine error-corrected qudit repeater. 

\begin{figure}[t]
  \centering
  \begin{minipage}{4em}
   \flushright $\frac{N_\mathrm{min}}{L[\mathrm{km}]}$
   \end{minipage}
  \begin{minipage}{26em}
  \includegraphics[width = \textwidth]{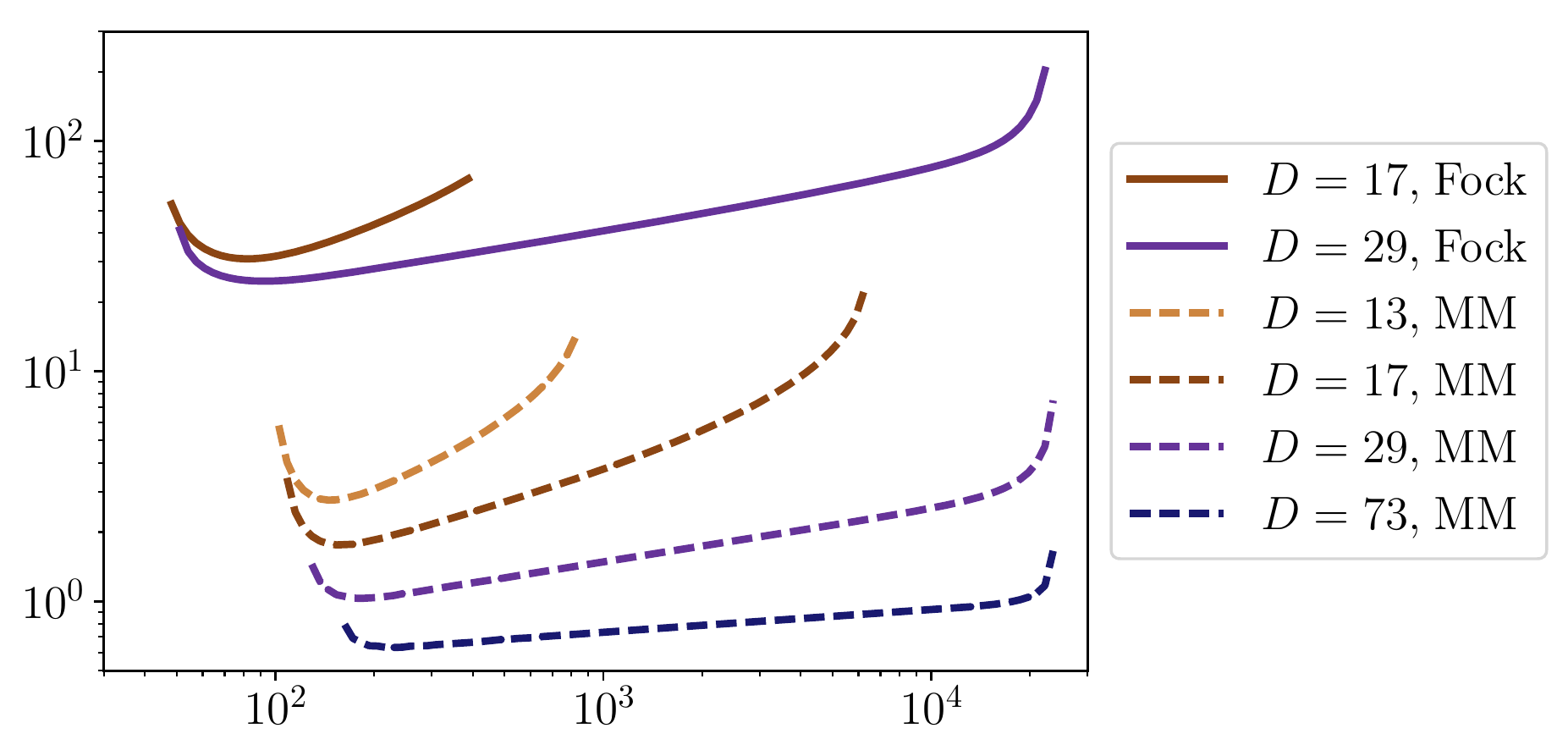}
  \end{minipage} 
   
   \hspace{-2.5em}
   $L$ in km
  \caption{The minimal number of repeater stations per km for which the PLOB-repeaterless bound can be just surpassed by a quantum repeater line of total length $L$ with MM and Fock qudits encoded by a $\llbracket D,1,(D+1)/2\rrbracket_D$ QECC. 
  The error parameters are $\alpha =0.2\mathrm{dB/km}$, $f_\mathrm{M}=10^{-2}$, $f_\mathrm{G}=10^{-3}$, and $\gamma = 10^{-2}\mathrm{dB/ms}$.
}
  \label{fig:how_many_stations}
\end{figure}

\begin{figure}[h] 
  \centering
  \begin{minipage}{4em}
   \flushright $H(P)$
   \end{minipage}
  \begin{minipage}{26em}
  \includegraphics[width = \textwidth]{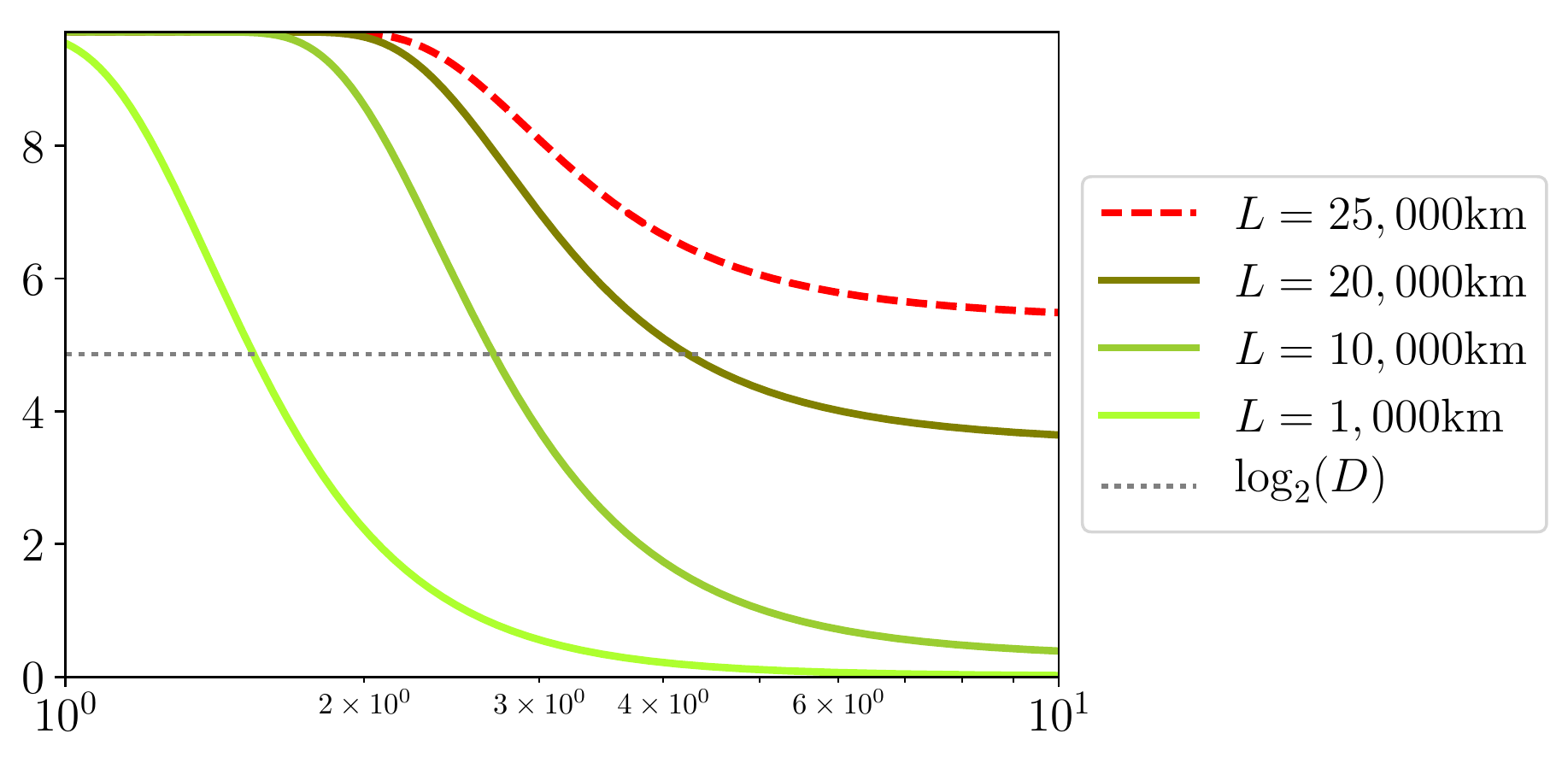}
  \end{minipage} 
  
   \hspace{-2.5em}
    $\frac{N}{L[\mathrm{km}]}$
  \caption{The Shannon entropy $H(P)$ of the error distribution of a state distributed by a $\llbracket29,1,15\rrbracket_{29}$ error-corrected  MM quantum repeater for different total lengths $L$ as a function of the inverse repeater spacing $1/L_0  = N/L$.
  Perfect error correction means $H(P)=0$. 
   For $L\le 20,000$km,  one can reach $H(P)\lesssim \log_2(D)$ by an adjustment of $N/L$. 
  Since the global minimum of $H(P)$ grows in $L$, 
  the value of  $N/L$ where $\log_2(D)$ intersects $H(P)$ exponentially increases.
  For  $L=25,000$km, $\lbrep = \max\{\log_2(D)- H(P),0 \}$ is zero for any choice of $N/L$, i.e., the quantum-repeater gain $\Delta$ is negative.
As in Fig.~\ref{fig:how_many_stations}, we use $\alpha =0.2\mathrm{dB/km}$, $f_\mathrm{M}=10^{-2}$, $f_\mathrm{G}=10^{-3}$, and $\gamma = 10^{-2}\mathrm{dB/ms}$.
  }
  \label{fig:entropy}
\end{figure}

A relevant figure of merit to compare different quantum repeater lines is the minimum number of repeater stations per unit length $N_\mathrm{min}/L$, which we plot in Fig.~\ref{fig:how_many_stations}
as a function of the total length $L$ for various encodings. 
All curves qualitatively show the same behavior:
For very small $L$, the PLOB-repeaterless bound is not surpassed, as direct quantum communication is still possible.
Eventually, with increasing $L$, the PLOB-repeaterless bound has dropped to a quantum capacity which can be surpassed by $\lbrep$.
At this minimal total length $L_\mathrm{min}$, the curves in Fig.~\ref{fig:how_many_stations} begin.
For Fock qudits, $L_\mathrm{min}$ is smaller than for MM qudits because fewer modes are used, consistent with previous observations above.
If $L$ is  slightly above $L_\mathrm{min}$, the PLOB-repeaterless bound  $\ubPLOB$ quickly approaches zero. 
Thus, the lower bound on the quantum capacity of the repeater, $\lbrep = \log_2(D) - H(P)$, is allowed to decrease as well, which leads to the possibility of setting up the quantum repeater with fewer repeater stations per unit length. 
This explains the initial drop of the curves for $L\gtrsim L_\mathrm{min}$. 
At some point, the global minima from Tab.~\ref{tab:how_many_stations} are reached.
For larger $L$, the regime $\ubPLOB \approx 0$ is entered. 
Since we consider $0 \lesssim \Delta$, this implies $H(P)\lesssim \log_2(D)$.
That is, the number of repeater stations is adjusted such that the Shannon entropy of the error distribution on the distributed state is kept slightly below $\log_2(D)$. 
If, in this regime,  
$L$ is increased, the amount of error correction overhead has to be adjusted accordingly.  
Therefore, the minimal number of repeater stations per unit length increases with $L$. 
The (log-log) slope of the corresponding curves in this intermediate region decreases with $D$ because QECCs with a larger code distances can more readily cope with the additional transmission losses.
For $D=73$ MM qudits, the code distance $d= 37$ of the QECC is large enough such that the 
curve just barely grows. 
Eventually, so many storage errors of Alice's quantum memory have accumulated that the  pseudothreshold\footnote{The (code capacity) pseudothreshold of a QECC is the error rate at which the physical error rate is equal to the logical error rate, in Eq.~\eqref{eq:p_succ_MM} with $f_\mathrm{S} = p_{e\neq0}$. 
For $\llbracket29,1,15\rrbracket_{29}$ and  $\llbracket73,1,37\rrbracket_{73}$  QECCs, our calculations show that this pseudothreshold is approximately $20\%$. By Eq.~\eqref{eq:storage_errors}, $f_\mathrm{S}=0.2$ is reached for $L\approx 20,000$km since  $\gamma = 10^{-2}\mathrm{dB/ms}$.} of the respective QECC is reached.
In that region, storage errors strongly influence $H(P)$ until the condition $H(P)\lesssim \log_2(D)$ cannot be fulfilled for any choice of $N/L$.
This explains the sudden growth of the curves in Fig.~\ref{fig:how_many_stations} for large values of $L$ and
is clearly visible in Fig.~\ref{fig:entropy}.

In conclusion of this subsection, we observe that for our error model and existing QECCs, 
genuine error-corrected qudit repeater lines require at least about $10^2$ and $10^3$ repeater stations for MM and Fock qudits, respectively.  
For a total length $L$ between $10^2$km and $10^4$km, the PLOB-repeaterless bound can be surpassed 
while the minimum number of repeater stations per unit length, $N_\mathrm{min}/L$ , gradually increases in $L$.
This increase is less pronounced for quantum repeaters with better error-correcting capabilities, i.e., for a higher qudit dimension $D$.

%% file: 4Conclusion.tex
 
In this paper, we have analyzed the applicability of error-corrected quantum repeaters based on higher-dimensional qudits as long-term candidates of a quantum communication infrastructure.
{To our knowledge, this work shows for the first time how the PLOB-repeaterless bound can be surpassed using error-corrected qudit repeaters.}
By making explicit how the PLOB-repeaterless bound relates to the encoding of abstract qudits into photonic modes, we obtain a bound on the capacity of the quantum repeater using the Shannon entropy of the error distribution of the final state.
We defined the quantum repeater-gain as a figure of merit and used it to identify genuine quantum repeaters by a systematic analysis of its dependency on a variety of parameters.

We derived an analytical solution of the quantum-repeater gain for error-corrected, one-way qudit repeaters based on two different types of physical encoding:
Fock encoding, where each qudit is encoded into a single photonic mode; 
and multimode encoding, where each computational basis state of a qudit has its own mode.
While Fock encoding is interesting from a theoretical perspective, as it allows to surpass the PLOB-repeaterless bound over shorter distances by harnessing higher photon numbers of the photonic mode, 
multimode qudits pose the more realistic way of implementing error-corrected qudit repeaters, as they are more  readily available in the form of e.g., time-bin qudits, temporal modes, and modes of orbital angular momentum. 
We found that genuine quantum repeaters are feasible if the distance $L_0$ between adjacent repeater stations is on the order of 1km for multimode encoding, independent of its total length. 
For Fock qudits, we can only prove that $L_0 \sim 10\mathrm{m}$ is sufficient however, we expect that this is due to our worst-case approximation of the pure-loss channel and that  Fock repeaters can also surpass the PLOB-repeaterless bound with $L_0 \sim 100\mathrm{m}-1\mathrm{km}$.

We have shown that an improvement of operational error rates makes it possible to decrease the necessary number of physical qudits per logical qudit, as well as the qudit dimension. 
For realistic error rates, the smallest qudit dimension with which a genuine quantum repeater could be realized within our error model is $D=13$ and $D=11$ for Fock and MM qudits, respectively.  
Although theoretical proposals for the generation of Fock states with an arbitrary photon number exist~\cite{BDSKW03}, high-quality Fock states have experimentally only been realized up to four photons, i.e., $D_\mathrm{Fock}\le 5$, and no significant improvement was made over the last 10-15 years~\cite{WDY06, TBHLNGS19}.
For MM qudits, on the other hand, the state-of-the-art continuously progresses: Qudits based on temporal modes, orbital angular momentum and time bin can be realized up to $D_\mathrm{TM}\le 7$~\cite{ADASBRTHS18}, $D_\mathrm{OAM}\lesssim 100$~\cite{KMEZ17}, and $D_\text{time-bin}\lesssim 10^{5}$~\cite{ZZHLVLRBMGNMSZWEWSW15, MMBVNMSG18}, respectively.
An experimental challenge that has to be overcome to realize the here-considered protocol is the development of high-fidelity, two-photon controlled-phase gates, as well as the preparation of multipartite entangled photons, in particular in the logical $\ket{+}$ state of a quantum polynomial code.

While near-term candidates such as the so-called single photon scheme based on a nitrogen vacancy architecture~\cite{RYGRHHWE18} and twin-field quantum key distribution~\cite{LYDS18, CAL18, GraCur19} are within experimental reach, they are spatially restricted to tens and hundreds of kilometers, respectively. 
As we showed here, error-corrected qudit repeaters, on the other hand, have the potential to overcome the PLOB-repeaterless bound over length scales on the same order of magnitude as the Trans-Siberian railroad, i.e., $10^{4}$km. 
These length scales are sufficient to connect any two points on earth.


Here, we have focused on subspace quantum error-correcting codes (QECCs)~\cite{QEC}, in particular, quantum polynomial codes~\cite{GL99, ABO08, KKKS06, Cross08}. 
However, recently it was shown  that subsystem QECCs can have an advantage in fighting leakage errors in ion trap quantum computers~\cite{BNB19}. It would be interesting to find out whether subsystem codes, such as Bacon-Shor codes~\cite{Shor95, Bacon06, AliCro07}, subsystem surface codes~\cite{BDCPS13}, 2D compass codes~\cite{LMNWB18}, and optimal generalized Bacon-Shor codes~\cite{Yoder19}, also have an advantage in coping with photon losses  in an error-corrected quantum repeater protocol.
{Finally, note that the Gottesman-Kitaev-Preskill (GKP) code
which encodes each physical qudit into a single bosonic mode is naturally suited in the case of leakage errors~\cite{GKP01}.
It would be valuable to extend our analysis for Fock qudits and multimode qudits to GKP qudits as well, similar to Ref.~\cite{VAWPT19}. 
}

\begin{acknowledgments}	
The authors thank
{Mohsen Razavi and }
Federico Grasselli 
for helpful discussions  
and Eric Sabo for feedback on the manuscript.
The authors acknowledge support from the Federal Ministry of Education and Research (BMBF, Project
Q.Link.X).
\end{acknowledgments}